\documentclass[aps,journal=prb,preprint]{revtex4-1}
\usepackage{graphicx}
\usepackage{epstopdf}
\usepackage{tabularx}
\usepackage{graphicx}

\begin{document}
\title{Origin of Stationary Domain Wall Enhanced Ferroelectric Susceptibility }
% \keywords{Keywords: ...}

\author{Shi Liu}
\email{sliu@carnegiescience.edu}
\affiliation{Extreme Materials Initiative, Geophysical Laboratory, Carnegie Institution for Science, Washington, D.C. 20015-1305 USA}
\author{R. E. Cohen}
\email{rcohen@carnegiescience.edu}
\affiliation{Extreme Materials Initiative, Geophysical Laboratory, Carnegie Institution for Science, Washington, D.C. 20015-1305 USA}
\affiliation{Department of Earth- and Environmental Sciences, Ludwig Maximilians Universit\"{a}t, Munich 80333, Germany}

\date{\today}
%\begin{document}
\begin{abstract}
%\abstract
{
Ferroelectrics usually adopt a multi-domain state with domain walls separating domains with polarization axes oriented differently. It has long been recognized that domain walls can dramatically impact the properties of ferroelectric materials. The enhancement of low-field susceptibility/permittivity under subswitching conditions is usually attributed to the reversible domain wall vibration. Recent experiments highlight the stationary domain wall contribution to the dielectric susceptibility irrespective of any lateral displacements or deformations of the wall. We study the effects of domain walls on low-field permittivity of PbTiO$_3$ with density functional theory and molecular dynamics simulations. The static dielectric constant is calculated as a function of increasing domain wall density and temperature. We find an increase of dielectric permittivity with increasing domain wall density, which is expected to occur at low driving field where the lateral motion of domain walls is forbidden. Real-space decomposition of dielectric response reveals that frustrated dipoles within the finite width of the domain walls are responsible for the enhanced low-field permittivity. 
}
\end{abstract}
\maketitle

Ferroelectrics characterized by the switchable spontaneous polarization under external field have served as critical components in electronics, optics, sensors, and actuators.~\cite{Scott07p954} Ferroelectrics  often possess complex domain structures with domain walls (DWs) separating homogeneously polarized domains. In response to an applied stimulus that favors one polarization state over another, the DW can move to increase the size of the favored domain.~\cite{Pramanick12p243} The susceptibility therefore consists of two contributions: the {\em intrinsic} contribution that originates from the polarization change within the bulk of the domains and the {\em extrinsic} contribution that arises from DW motions. It is now widely recognized that DWs can have a profound effect on the dielectric, piezoelectric and pyroelectric susceptibilities of ferroelectric materials.~\cite{Zhang94p454,Taylor97p1973,Xu01p1336,Xu14p3120,Karthik11p024102,Karthik12p167601,Zubko16p524} For example, experiments suggested that most ($> 60$\%) of the dielectric and piezoelectric responses at room temperature in lead zirconate-titanate (PZT) ceramics is from the DW contributions~\cite{Zhang94p454,Xu01p1336}, with both 180$^\circ$ and non-180$^\circ$ walls contributing to dielectric response and non-180$^\circ$ walls affecting piezoelectricity.~\cite{Pramanick11p293,Zednik11p3104}

Controllably optimizing the susceptibilities of ferroelectrics through DW engineering requires a microscopic understanding of the dynamics of DWs in response to a stimulus. For a driving field of large amplitude, the DW contribution to the susceptibility shows strong field-amplitude dependence, which is attributed to the field-induced irreversible DW motion. The upper bound of the dielectric permittivity due to the displacement of 180$^\circ$ DW can be approximated as $\varepsilon^{\rm DW} \cong 2P_s/\varepsilon_0E_c$, where $P_s$ is the bulk polarization, $E_c$ is the coercive field and the factor of two comes from polarization reversal. For BaTiO$_3$, using $P_s=0.25$ C/m$^2$ and $E_c = 1$kV/cm gives $\varepsilon^{\rm DW}  \cong 560,000$. The total dielectric response is the weighted average of DW contribution and intrinsic bulk contribution, and is often much smaller than the upper bound due to the low volume fraction of DW. 

For a stimulus that is much smaller than the coercive field, a number of mechanisms have been proposed to explain DW contrition to the enhanced dielectric response in the absence of DW motion.~\cite{Lawless70p419,Marvan69p482} Lawless and Fousek proposed that the DW has excessive polarizability because the materials within the wall have polarization passing through zero and can be considered as being closer to the phase transition than materials in the bulk.~\cite{Lawless70p419} The temperature gradient and heat transfer across DWs induced by the electrocaloric effect of antiparallel domains was also suggested to affect the frequency dispersion of dielectric susceptibility.~\cite{Marvan69p482} 

The idea of reversible DW vibration was also proposed to explain small-signal response. Unlike the typical DW motion that moves from one Peierls potential to another via nucleation-and-growth mechanism,~\cite{Liu16p360} reversible DW motion comes from DW displacement inside one minimum of the Peierls potential.~\cite{Tagantsev10Book} The DW is considered as an oscillator with some effective mass and vibrates around the equilibrium position with displacement amplitude determined by electric and/or elastic restoring forces.~\cite{Kittel51p458,Fousek64p830,Arlt87p37,Arlt91p2283,Pertsev95p135,Pertsev96p1364} However, the estimated DW displacement is only few percent of a lattice constant~\cite{Lawless70p419,Arlt87p37}, which is not well defined microscopically considering that 1) atoms are discreet in crystalline solids; 2) a DW is at least one unit cell wide; 3) atoms move {\em perpendicular} to the direction of the DW vibration. Furthermore, the assumption that a DW has effective mass directly implies that a DW would exhibit inertial response, which was a subject of debate with both confirming~\cite{Kim10p1266,Dawber03p436} and contradicting results~\cite{Sharma13p1323,Molotskii07p271}. Molecular dynamics simulations of post-field DW behavior show that ferroelectric DWs have no significant intrinsic inertial response.~\cite{Liu13p232907} All these results challenge the concept of reversible DW vibrations at low field. Recent experiments in (111)-oriented PbZr$_{0.2}$Ti$_{0.8}$O$_3$ thin films that effectively freeze out DW motions indicate that the stationary contribution from the DW can be 6-78 times larger than bulk response~\cite{Karthik12p167601,Xu14p3120}, but the exact microscopic nature of stationary DW contribution remains unclear. 

As a canonical ferroelectric oxide, PbTiO$_3$ and DWs in PbTiO$_3$ have been the subject of numerous first-principles studies.~\cite{Meyer02p104111,Poykko99p2830,Wojde14p247603} The 180$^\circ$ DWs are usually modeled with a $Na\times1a\times1c$ supercell ($a$ and $c$ are short-axis and long-axis lattice constants) where the unit cells are stacked in the $x$ direction and $N/2$ unit cells have polarization aligned along $+z$ while $N/2$ unit cells have polarization aligned along $-z$. Recent first-principles studies pointed out that the DW may possess a net polarization along $+y$ direction, which could undergo a temperature-driven ferroelectric-paraelectric transition confined to the DW.~\cite{Wojde14p247603} It is well known that the theoretical lattice constants of ferroelectrics depend sensitively on the density functional approximations, with local density approximation (LDA) underestimating the tetragonality ($c/a$) while generalized gradient approximation (GGA) yielding a wrong supertetragonal structure.~\cite{Rabe07Book} Even using the same density functional, the first-principles structural parameters for the fully-optimized tetragonal PbTiO$_3$ reported in literatures~\cite{Bilc08p165107,Sharma14p214102} vary slightly, partly because of the different types of pseudopotentials (PPs, e.g., projector-augmented wave method, norm-conserving, and ultrasoft) employed in the calculations. Because the ferroelectric instabilities are extremely sensitive to the lattice constants, we first benchmark the performances of several first-principles methods using three density functional approximations, LDA, Wu-Cohen (WC),~\cite{Wu06p235116} and PBEsol,~\cite{Zhao08p184109} and three open-source PP libraries, GBRV ultrasoft pseudopotential (USPP) library,~\cite{Garrity14p446} PSlibrary(v1.0.0),~\cite{DalCorso14p337} and Bennett-Rappe normconersing (NC) PP library.~\cite{Rappe90p1227} All calculations are carried out using Quantum Espresso~\cite{Giannozzi09p395502} with a $4\times4\times4$ Monkhorst-Pack grid and a force convergence threshold of 1.0$\times 10^{-4}$ Ry/Bohr and energy convergence threshold of 1.0$\times 10^{-5}$ Ry. Our results (Table I) for the fully-relaxed tetragonal PbTiO$_3$ are consistent with previous theoretical investigations.~\cite{Bilc08p165107,Sharma14p214102,Wu06p235116,Zhao08p184109}  

Because the first-principles lattice constants deviate from the experimental values, there are three possible supercell models for constructing 180$^\circ$ DWs: I) fix the lattice constants to the experimental values and relax internal atomic coordinates; II) fix the lattice constants to the theoretical values of the fully-optimized tetragonal unit cell and relax the atomic positions; III) fully relax the lattice constants and atomic positions of the suprecell. We carry out benchmark calculations to study the effects of computational setups on the structural properties and energetics of 180$^\circ$ DWs uisng a $12a\times1a\times1c$ supercell. The structures are optimized with a $1\times4\times4$ Monkhorst-Pack grid and the same convergence thresholds as for unit cell benchmark studies. The DW energy ($E_{\rm DW}$) is calculated with $E_{\rm DW} = \frac{1}{2S}\left(E_{\rm supercell} - E_{\rm SD}\right)$, where $S$ is the DW area, $E_{\rm supercell}$ is the energy of the supercell with two DWs, and $E_{\rm SD}$ is the energy of the single-domain supercell of the same lattice constants. We find that the DW structure and energy depend on both the density functional approximation and the supercell model (Table II). For Model I, the three density functionals, LDA, WC, and PBEsol, predict similar DW energies while LDA gives the highest value. This is likely due to the more severe artificial inhomogeneous tensile strain effect of LDA when using experimental lattice constants because LDA underestimated the lattice constants. For a given density functional, fully relaxing the lattice constants of the supercell (Model III) leads to a $c/a$ ratio smaller than the ground state bulk value. This suggests that 180$^\circ$ DWs, though usually considered to be non-ferroeleastic, can give rise to mechanical clamping that suppresses the tetragonality of nearby unit cells. However, such mechanical clamping is likely to be only important for domain structures with high DW density and small DW-DW distance. 

\begin{figure}[t]
\centering
\includegraphics[scale=1.0]{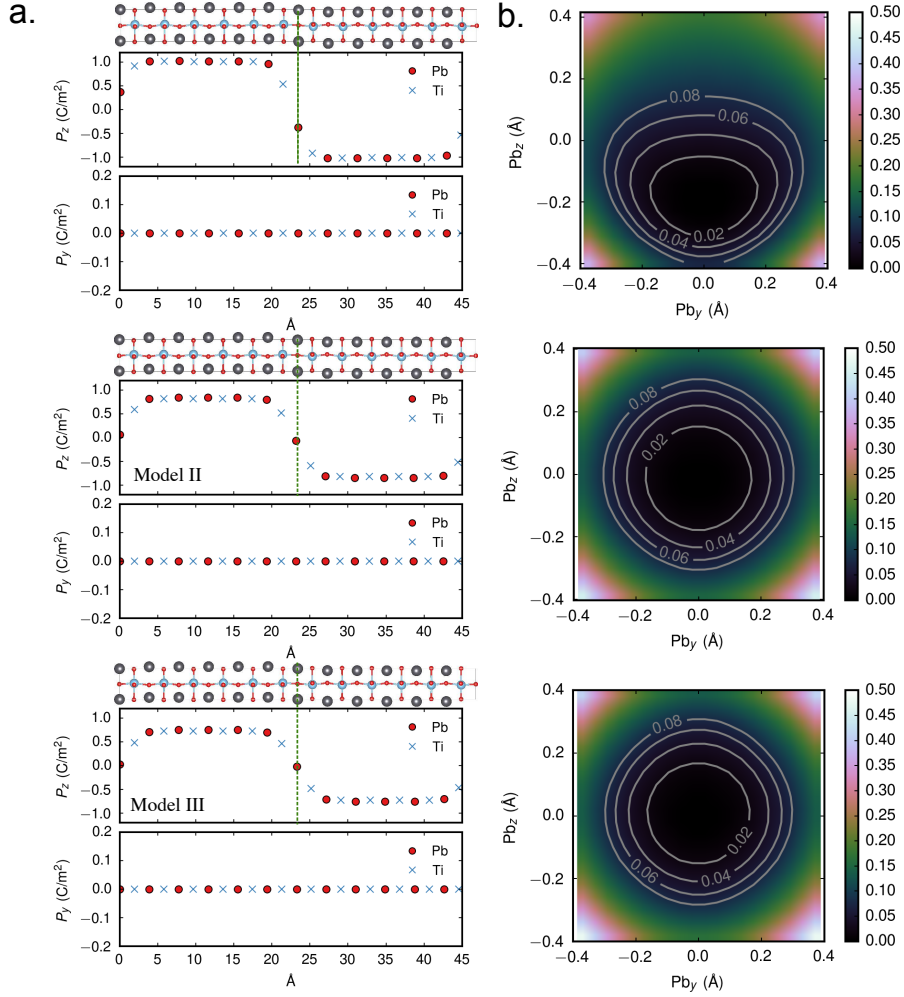}\\
 \caption{First-principles modeling of 180$^\circ$ DWs in PbTiO$_3$ using different supercell models and LDA. (a) Polarization profiles for structures with 180$^\circ$ DWs; (b) Two-dimentional potential energy surfaces for Pb atoms at DWs (highlighted with dashed green lines in pannel a).}
 \label{compare}
 \end{figure}

The examination of the polarization profile across the domain boundary (FIG.~\ref{compare}a) reveals some subtle differences between supercell models. All three models show that the DW is essentially centered at the PbO plane, as demonstrated by the Ti-centered local polarization changing from $+z$ to $-z$ across the PbO plane, and the DW does not have net polarization along the $y$ axis. However, for Model I, the Pb-centered local polarization at the DW is not exactly zero along the $z$ axis, different from those in Model II and III. This is again due to the artificial strain effect of LDA for using experimental lattice constants. We also find that for the two DWs in the supercell, they could have antiparallel polarization along the $z$ axis (one DW has $P_z$ $>$ 0 and the other DW has $P_z$ $<$ 0) and parallel polarization (both DWs have polarization aligned along $+z$ or$ -z$). These two configurations are close in energy ( $<$ 5 meV/supercell with LDA), with the configuration of antiparallel DW polarization slightly lower in energy. 

\begin{figure}[t]
\centering
\includegraphics[scale=1.0]{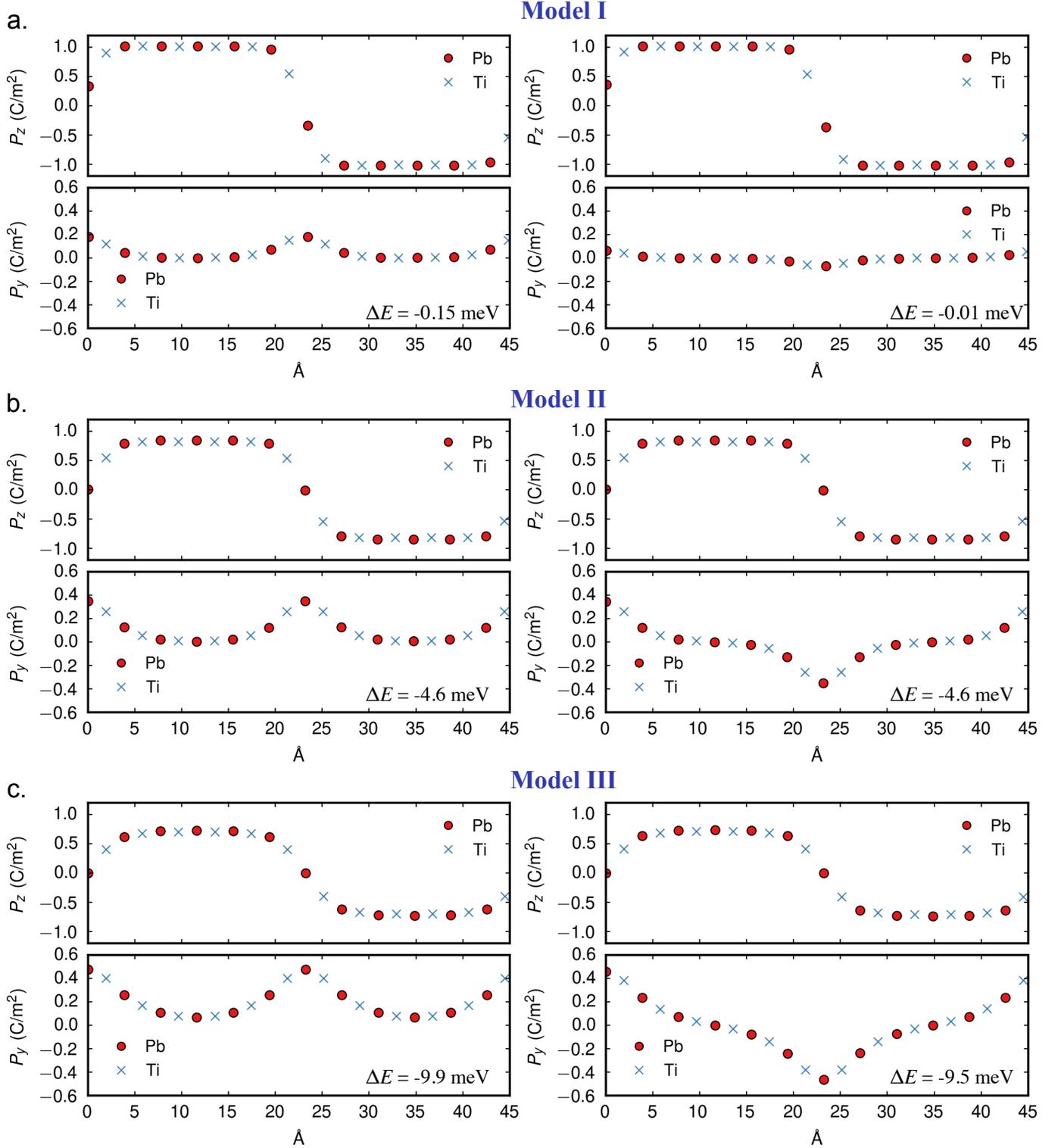}\\
 \caption{Polarization profiles for mutldomain configurations with parallel and antiparallel domain wall polarization along the y axis ($P_y^{\rm DW}$) obtained using three supercell models and LDA functional. The energy difference per DW ($\Delta E$) is calculated with respect to the energy of the configuration with $P_y^{\rm DW}$ = 0 C/m$^2$.  }
  \label{compare2} 
 \end{figure}

We calculate the two-dimensional potential energy surface (PES) for the Pb atom at the DW with LDA. It is found that the local PES confined to the DW (FIG.~\ref{compare}b) is extremely shallow along the $y$ axis, regardless the supercell model employed. The Pb atom can easily be displaced away from their equilibrium position with little energy cost. By starting from a configuration with Pb atoms at DWs displaced slightly along the $y$ axis, we obtained two more optimized configurations (FIG.~\ref{compare2}) for each supercell model: one configuration has parallel DW polarization and the other configuration has antiparallel DW polarization along the $y$ axis. The three configurations (one without DW polarization and two with DW polarization along $y$) are close in energy (energy difference smaller than 0.01 eV/supercell), confirming the shallow PES at the DW. The calculations of WC and PBEsol give essentially the same results, both functionals predicting three configurations with small energy difference. This indicates that atoms within DWs are easier to move under a driving field. Note that the type of supercell model and the density functional approximation can influence subtly the magnitude and directions of local polarization confined to the DW, it remains an open question about the robustness of the ferroelectric transitions at DWs suggested in previous first-principles studies ~\cite{Wojde14p247603} under experimental conditions. However, the presence of polarization suppression along the $z$ axis at the DW and the shallow local PES appear to be robust, independent of computational setups. 

\begin{figure}[t]
\centering
\includegraphics[scale=1.0]{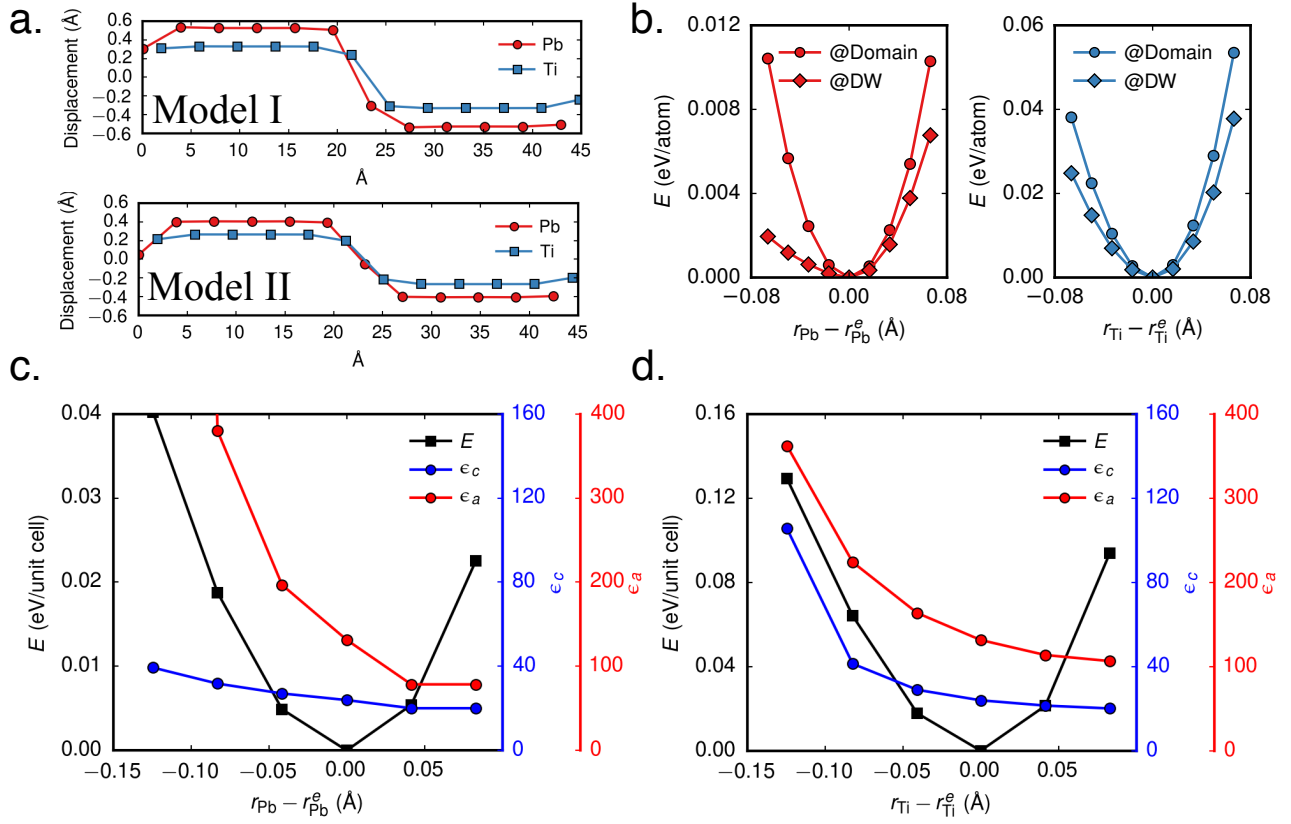}\\
 \caption{(a) $z$ displacement of Ti and Pb atoms relative to the center of oxygen cage. (b) Potential energy surfaces for Pb and Ti atoms at DWs (@DW) and in the bulk (@Domain). Energy and dielectric constants of 5-atom unit cell of PbTiO$_3$ as a function of atomic distortion along the polar axis for Pb (c) and Ti (d). }
  \label{DFTDie}
 \end{figure}

The smaller polarization at DWs is due to the smaller Pb and Ti atomic displacements with respect to the center of oxygen cage (FIG.~\ref{DFTDie}a). Accordingly, Pb and Ti atoms at DWs (@DW in FIG.~\ref{DFTDie}b) have much shallower potential energy surfaces for small distortions away from the equilibrium position. We investigate the effect of atomic displacement on the value of static dielectric constant of PbTiO$_3$ with density perturbation functional theory (DPFT) using the ABINIT code.~\cite{Gonze09p2582} We start with a fully relaxed 5-atom unit cell of PbTiO$_3$ and apply small distortions to the Ti (Pb) atom. For each distorted structure, we calculate the short-axis ($\varepsilon_a$) and long-axis ($\varepsilon_c$) dielectric constants. It is found that structures with suppressed Pb and Ti displacement supercell have both larger $\varepsilon_a$ and $\varepsilon_c$ (FIG.~\ref{DFTDie}c-d). This shows qualitatively that the DW characterized with smaller atomic displacements will possess higher dielectric susceptibility. 

To understand the finite-temperature dielectric response of DWs, we perform all-atom molecular dynamics (MD) simulations taking the ferroelectric 180$^\circ$ DW and ferroelastic 90$^\circ$ DW in PbTiO$_3$ as examples. Combined with accurate interatomic potentials derived from {\em ab initio} calculations~\cite{Sepliarsky02p36,Sepliarsky11p435902,Liu13p102202,Liu13p102202}, MD simulations have been applied to study various aspects of ferroelectrics~\cite{Liu13p232907,Rose12p187604,Zeng11p142902,Hashimoto14p1074} in different environments.~\cite{Sepliarsky05p014110,Xu15p79,Takenaka13p147602}  Our force field of PbTiO$_3$ is developed based on the bond-valence theory and is parametrized from first-principles.~\cite{Shin05p054104,Liu13p102202} All simulations are performed under constant-volume constant-temperature ($NVT$) conditions over a wide range of temperatures (150--320~K) with lattice constants of unit cells fixed to experimental values. To determine the effect of DW density on the total dielectric response, the simulations for 180$^\circ$ DWs are carried out over a 48$a\times$8$a\times$8$c$  perovskite-type supercell with varying numbers (2, 4, 6, 8, 10, 12) of walls under periodic boundary conditions. An 80$a_x\times$6$a_y\times$80$a_z$ supercell is used to model domain structures with 90$^\circ$ walls, where $a_x = (a+c)/2$, $a_y=a$, and $a_z=(a+c)/2$ are averaged lattice constants along Cartesian axes. This choice of supercell dimensions leads to a typical $c/a/c/a$ multidomain structure.~\cite{Pertsev96p6401} 

The local static dielectric permittivity (susceptibility) tensor at unit cell $m$, $\varepsilon_{ij}^m$ ($\chi_{ij}^m$), is calculated using fluctuation formulas~\cite{Wemple69p547,Boresch00p8743,Ponomareva07p227601,Ponomareva07p235403, Ponomareva08p012102,Hashimoto14p1074}:
\begin{equation}
\chi_{ij}^m \approx \varepsilon_{ij}^m=\frac{V }{\varepsilon_0k_BT}\left(\left< P_i^mP_j^m\right> - \left<{P_i}^m\right> \left< {P_j}^m\right>\right)
\end{equation}
where $\varepsilon_0$ is vacuum permittivity, $k_B$ is the Boltzmann constant, $T$ is the temperature, $V$ is the volume of unit cell, $i$ and $j$ define Cartesian components, 
$P_i^m$ is the local (within the unit cell $m$) polarization of {\em i}th component, and $\left<...\right>$ represents the thermal average.  The magnitude of polarization fluctuations dictates the magnitude of dielectric response. The instantaneous local polarization $\mathbf{P}^m(t)$ is
\begin{equation}
\mathbf{P}^m(t) = \frac{1}{V}\left(\frac{1}{8}\mathbf{Z}^*_{\rm Pb}\sum_{i=1}^{8}\mathbf{r}_{{\rm Pb},i}^m(t)  +\mathbf{Z}^*_{\rm Ti}\mathbf{r}_{{\rm Ti}}^m(t) + \frac{1}{2}\mathbf{Z}^*_{\rm O}\sum_{i=1}^{6}\mathbf{r}_{{\rm O},i}^m(t) \right)
\end{equation}
where $\mathbf{Z}^*_{\rm Pb}$, $\mathbf{Z}^*_{\rm Ti}$, and $\mathbf{Z}^*_{\rm O}$ are the Born effective charges of Pb, Ti, and O atoms; $\mathbf{r}_{{\rm  Pb},i}^m(t)$, $\mathbf{r}_{{\rm Ti},i}^m(t)$, and $\mathbf{r}_{{\rm O},i}^m(t)$ are instantaneous atomic positions in unit cell $m$. The validity of equation (1) has been confirmed numerically for both homogeneous and nonhomogeneous nanostructures of different dimensionalities,~\cite{Ponomareva07p227601,Ponomareva07p235403} and it allows convenient and accurate estimation of dielectric responses without directly applying electric field, ideal for studying low-field response. The total dielectric response of the supercell is estimated by taking the average of the local dielectric permittivity or local dielectric susceptibility ($1/\varepsilon_{ij}^m$), depending on the connection style (parallel vs. serial) of unit cells (approximated as capacitors).

\begin{figure}[t]
\centering
\includegraphics[scale=2.0]{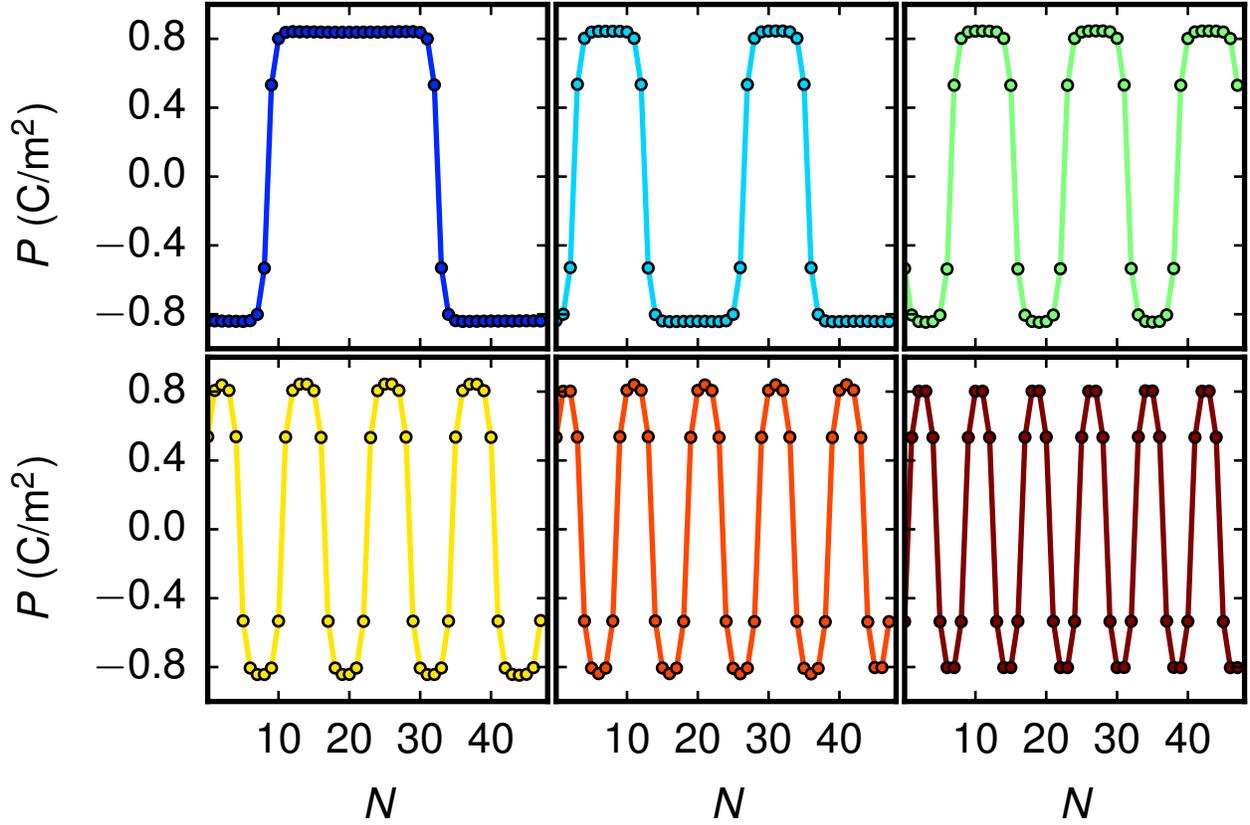}\\
 \caption{Simulated layer-resolved polarization profiles for supercells with 2, 4, 6, 8, 10, and 12 DWs with a $48\times8\times8$ supercell. The layer in $y$--$z$ plane is indexed along the $x$ axis. The layers at domain boundaries have smaller polarizations.}
   \label{MD180DW}
 \end{figure}

\begin{figure}[t]
\centering
\includegraphics[scale=2.0]{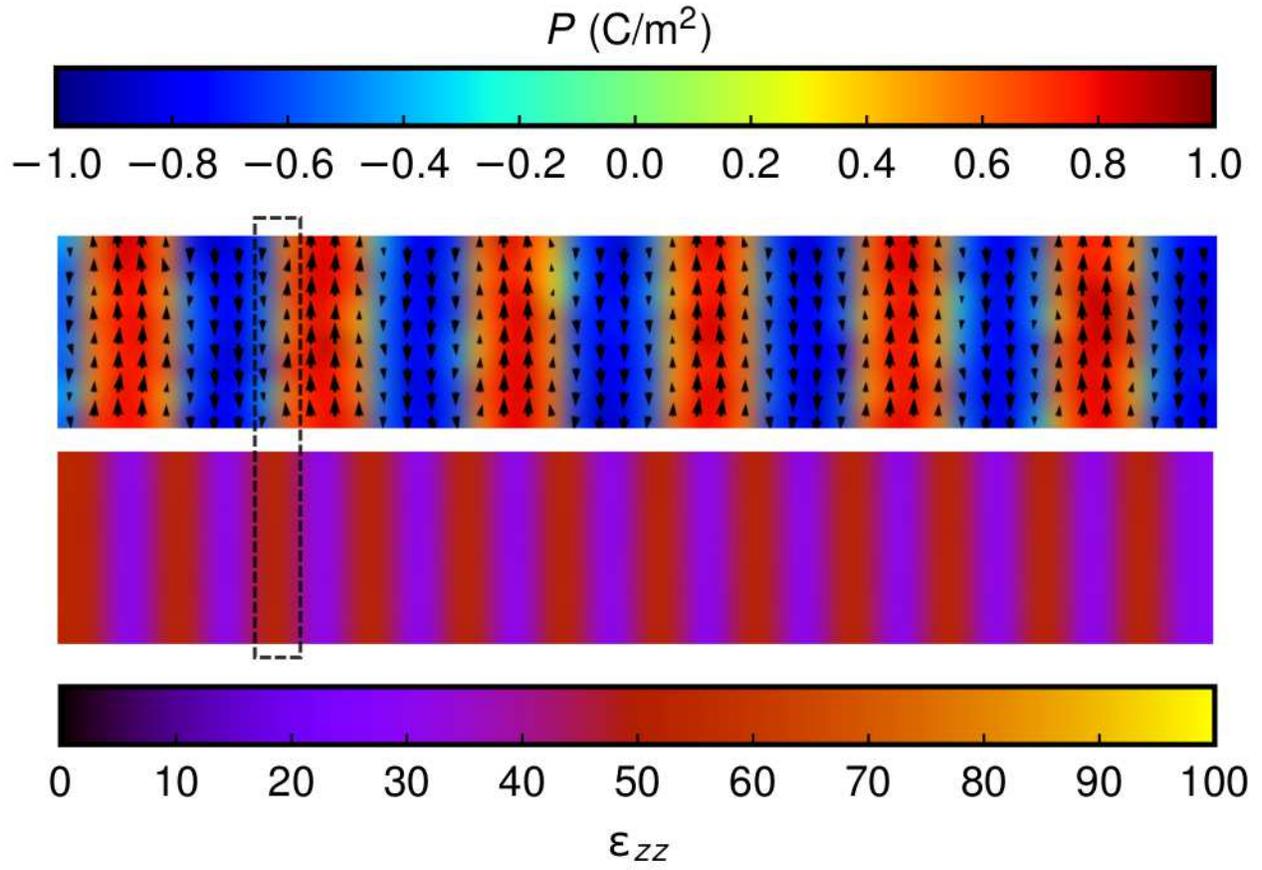}\\
 \caption{Profiles for local polarization $P^m_z$ (top) and local $\varepsilon_{zz}^m$ (bottom). The black arrows represent local dipoles. The position of one DW is highlight by the dashed rectangle.}
  \label{MD180DWDie}
 \end{figure}

The polarization profiles for domain structures with 180$^\circ$ walls at 300~K obtained from MD simulations reveal frustrated polarization at domain boundaries (FIG.\ref{MD180DW}). The width of a DW is approximately two unit cells thick, consistent with first-principle calculations. Within DWs, the polarization ($P^{\rm DW}\approx 0.45$~C/m$^2$) is much smaller than the polarization in adjacent bulk-like domains ($P^{\rm bulk}\approx 0.8$~C/m$^2$). This is due to the competition between minimizing the local energy, which favors small deviation of local polarization from the bulk value, and minimizing the gradient energy, which prefers a small polarization change, $2P^{\rm DW}$, across the domain boundary. The dipoles of smaller magnitude are expected to be more susceptible to external stimuli. We consider the dielectric permittivity in response to an electric field applied along the polar axis ($z$ axis), $\varepsilon_{zz}$. Figure \ref{MD180DWDie} presents the unit-cell-resolved polarization profile and dielectric permittivity for a supercell with 12 walls. The layers within the DWs clearly possess higher permittivity ($\approx 60$), and the layers in the domains have permittivity comparable to single domain value ($\approx 30$). This is a direct consequence of larger polarization fluctuation associated with smaller dipoles. 

\begin{figure}[t]
\centering
\includegraphics[scale=2]{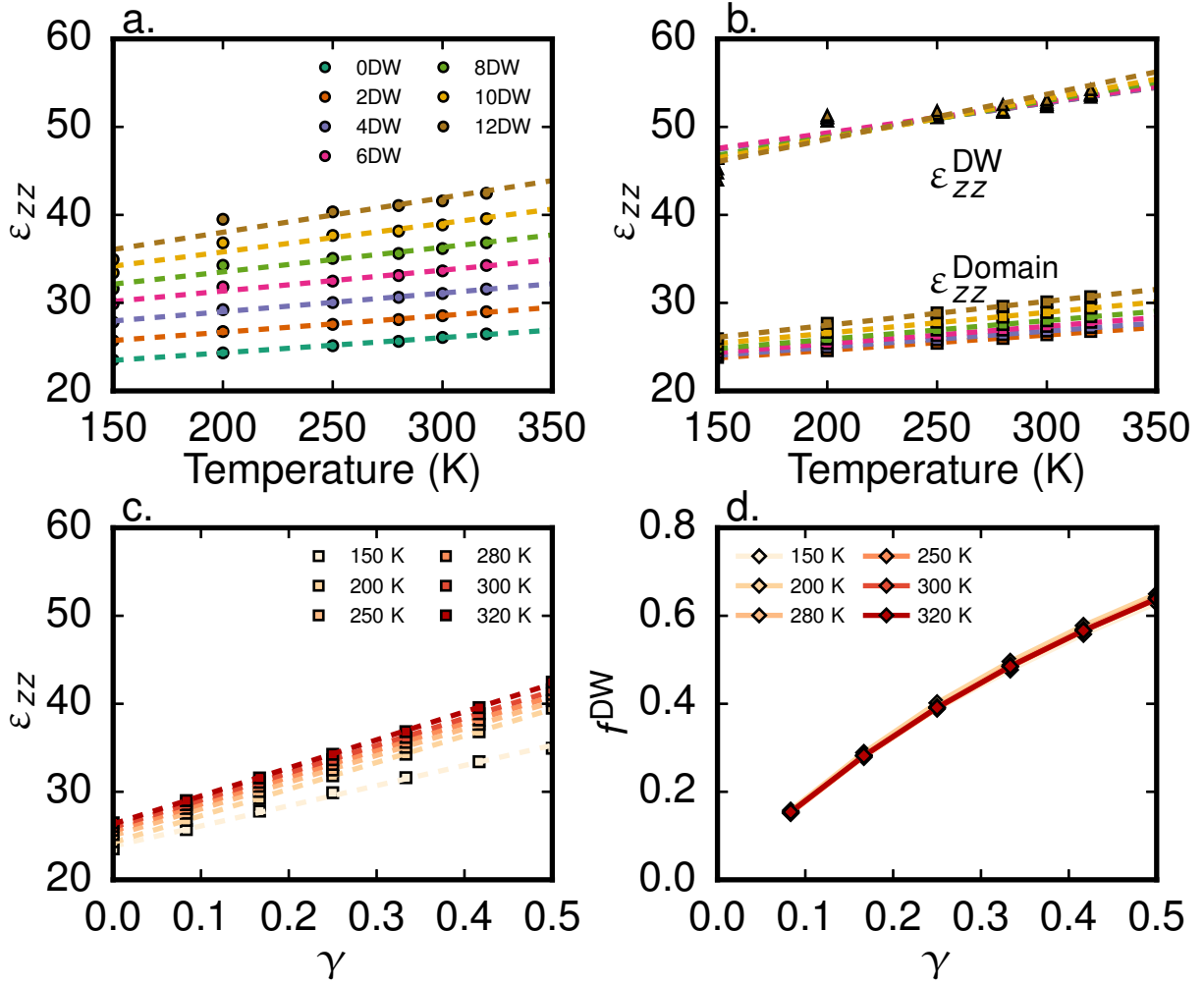}\\
 \caption{Temperature dependence of (a) total dielectric constant $\varepsilon_{zz}$ and (b) DW dielectric constant $\varepsilon_{zz}^{\rm DW}$ and domain dielectric constant $\varepsilon_{zz}^{\rm D}$ for supercells containing different number of 180$^\circ$ walls. DW volume fraction ($\gamma$) dependence of (c) total dielectric constant $\varepsilon_{zz}$ under different temperatures and (d) the weight of DW contribution ($f^{\rm DW}$). }
  \label{MD180DWDie2}
 \end{figure}

We further estimated the total dielectric constant as a function of temperature and the volume fraction ($\gamma$) of DWs. The volume fraction of DWs is defined as $2n^{\rm DW}/N$ where $n^{\rm DW}$ is the number of DWs, $N=48$ is the supercell dimension along the $x$ axis, and the factor of two comes from the observation that each DW consists of two layers of unit cells. We find a nearly linear dependence of $\varepsilon_{zz}$ on temperature below 350~K, far below the phase transition temperature of PbTiO$_3$ (FIG.\ref{MD180DWDie2}a), and $\varepsilon_{zz}$ increases almost linearly with $\gamma$ under a given temperature (FIG.\ref{MD180DWDie2}c). The real-space decomposition of the total $\varepsilon_{zz}$ allows the examination of individual contributions from DWs ($\varepsilon_{zz} ^{\rm DW}$) and domains ($\varepsilon_{zz} ^{\rm D}$), with the toal $\varepsilon_{zz} = \gamma \varepsilon_{zz} ^{\rm DW} + (1 - \gamma) \varepsilon_{zz} ^{\rm D}$ as the multidomain configuration can be viewed as a set of parallel capacitors for the response along $z$. $\varepsilon_{zz} ^{\rm DW}$ and $\varepsilon_{zz} ^{\rm D}$ have different temperature dependences (FIG.\ref{MD180DWDie2}b), with $\varepsilon_{zz} ^{\rm DW}$ possessing a stronger  temperature dependence (larger $\varepsilon_{zz}^{\rm DW}$--$T$ slope) than $\varepsilon_{zz}^{\rm D}$ . Interestingly, a noticeable increase in $\varepsilon_{zz}^{\rm D}$ with decreasing domain size is observed. In addition, the contribution of DWs to the total dielectric response, defined as $f^{\rm DW} = \gamma \varepsilon_{zz} ^{\rm DW}/\varepsilon_{zz}$, is as high as 50\% for a DW volume fraction of 0.35 (FIG.~\ref{MD180DWDie2}d). These results demonstrate that {\em even in the absence of DW motion}, the stationary 180$^\circ$ DW structure itself can enhance the dielectric response by almost a factor of two at room temperature. 

\begin{figure}
\centering
\includegraphics[scale=2]{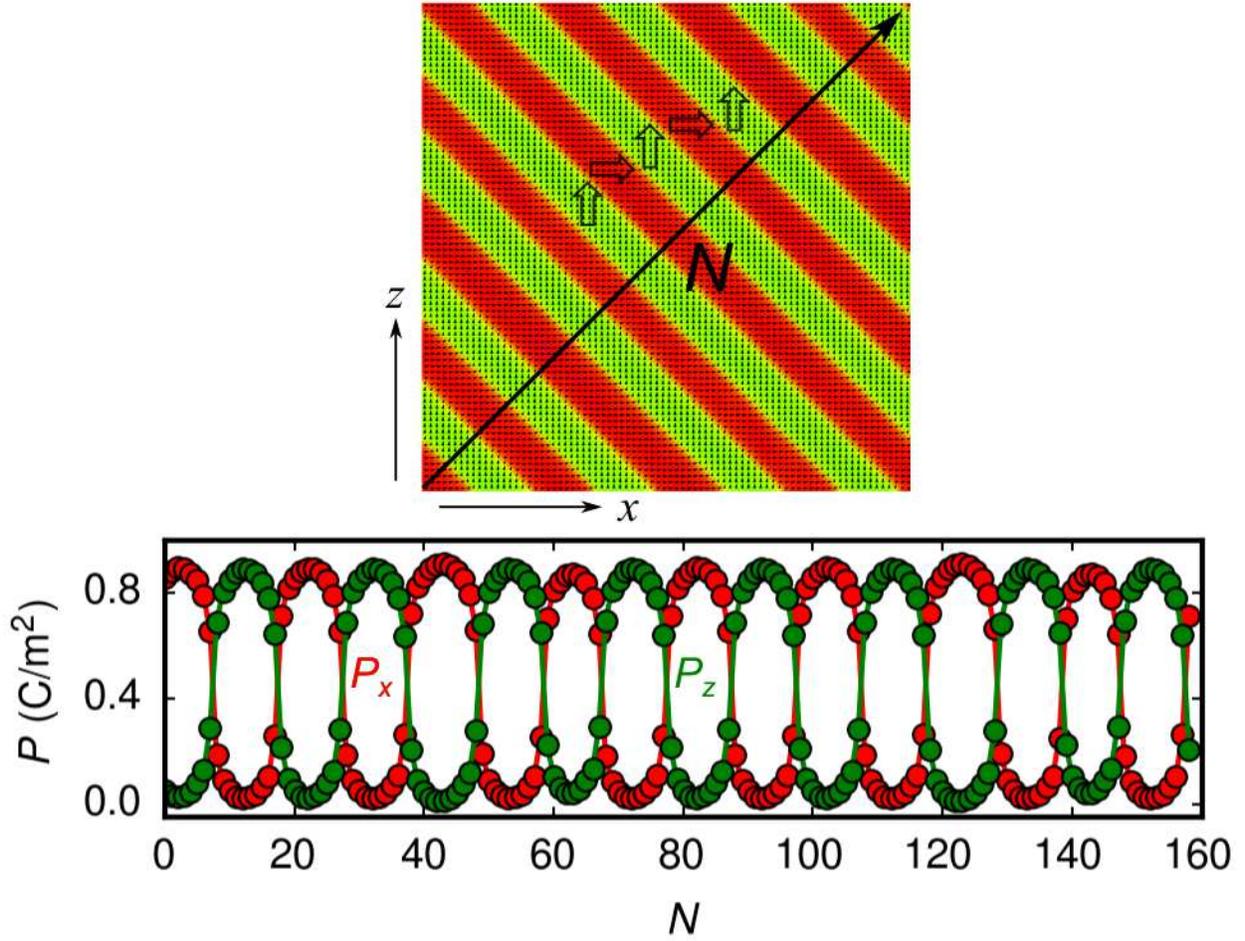}\\
 \caption{Molecular dynamics simulations of 90$^\circ$ DWs. The zig-zag domain pattern (top) with 90$^\circ$ DWs separating $P_x$ (red) domains and $P_z$ (green) domains. Layer-resolved $x$-component and $z$-component polarization profile (bottom) for supercells with 16 walls. The layer is indexed along [110] direction.}
 \label{MD90DW}
 \end{figure}

Similar susceptibility enhancement applies to 90$^\circ$ walls as well. Figure \ref{MD90DW} shows the simulated layer-resolved polarization profiles at 300 K for multidomain structures with 90$^\circ$ walls separating $+P_z$ (green) and $+P_x$ (red) domains. The width of 90$^\circ$ DWs is $4-5$ unit cells, and dipoles within DWs have smaller magnitudes, just as 180$^\circ$ DWs. We used an orthorhombic supercell, so the angle between the polarization axes of neighboring domains is exactly 90$^\circ$ instead of 2arctan($a/c$) that is geometrically required for a tetragonal ferroelectric. The domains are therefore strained. We find that when the domain size is comparable to the DW thickness, the unit cells inside domains also obtain smaller polarization. 

\begin{figure}
\centering
\includegraphics[scale=1]{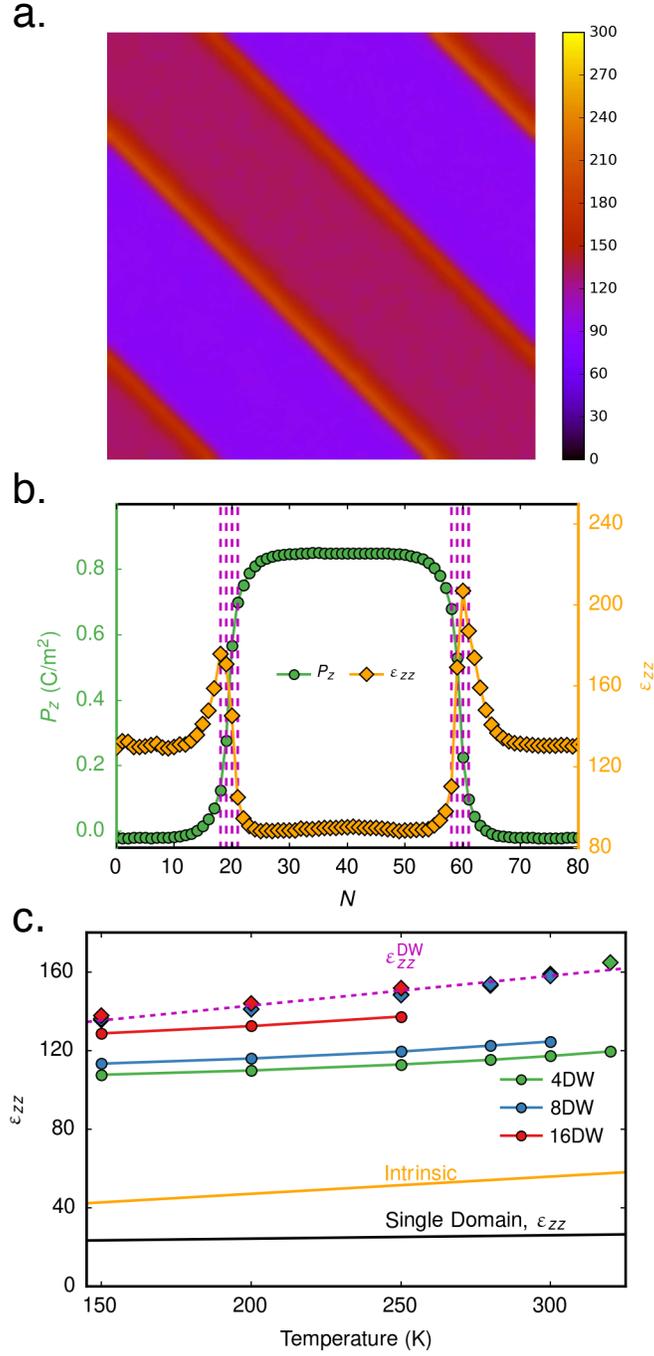}\\
 \caption{(a) Profile of local $\varepsilon_{zz}^{\rm DW}$ for a supercell with four 90$^\circ$ walls at 300~K. (b) Layer-resolved polarization profiles and $\varepsilon_{zz}$. Only half of the supercell is shown. Temperature dependence of (c) total $\varepsilon_{zz}$ (circle) and (d) DW dielectric constant $\varepsilon_{zz}^{\rm DW}$ (square) for supercells containing different numbers of 90$^\circ$ walls. }  \label{MD90DWDie}
 
 \end{figure}

We focus on the $\varepsilon_{zz}$ component. The real-space profile of local $\varepsilon_{zz}^m$ for a supercell with four walls at 300~K reveals that domain boundaries have larger permittivity values (FIG.~\ref{MD90DWDie}a). Layer-resolved $\varepsilon_{zz}$ (FIG.~\ref{MD90DWDie}b) provides more details. As expected, the $\varepsilon_{zz}$ of $P_z$ domains is smaller than that of $P_x$ domains. This agrees with first-principles~\cite{Yu09p165432,Pilania12p7580} and experimental~\cite{Foster93p10160,Fontana91p8695} results that PbTiO$_3$ in the tetragonal phase has a higher dielectric constant along the short axis. The values of local permittivity at 90$^\circ$ DWs are about two times higher than those in domains. We also note that the $\varepsilon_{zz}$ of $P_z$ domains in domain structures with only 90$^\circ$ walls is higher than that in domain structures with only 180$^\circ$ walls (FIG.~\ref{MD180DWDie}). This comes from the strain effect due to the usage of an orthogonal supercell as discussed above, and is consistent with the previous studies that strained ferroelectric thin films and superlattices have larger dielectric constants,~\cite{Bungaro04p184101,Park01p533} and is also consistent with our DFT calculations that reduced atomic displacements cause higher dielectric response (FIG.~\ref{DFTDie}c-d). The intrinsic $\varepsilon_{zz}$ for a domain structure with 50\% $a$ domain and 50\% $c$ domain is about 60, estimated with $2/(\varepsilon_a^{-1} + \varepsilon_c^{-1}$) assuming $c$ and $a$ domains as serial capacitors. Our simulations show that the domain structure with 90$^\circ$ walls has much larger dielectric response, and more DWs lead to higher values (FIG.~\ref{MD90DWDie}c). This highlights the contribution form stationary DWs. The $\varepsilon_{zz}^{\rm DW}$ of 90$^\circ$ wall in PbTiO$_3$ is about 6 times larger than the single domain value. It is noted that for the supercell with 16 walls, we observe domain merger at temperatures above 250~K due to thermal broadening of DWs: the final domain structure contains only four 90$^\circ$ walls. 

The enhanced dielectric constants resulting from smaller dipoles at domain boundaries also have important indications for the nucleation step in ferroelectric switching. The classic Miller-Weinreich (MW) nucleation model~\cite{Miller60p1460} for DW motion assumes depolarization interactions between boundary charges dominate the nucleation step. To reduce the repulsive energy penalty due to Coulomb repulsions between positive depolarization charges, the MW model incorrectly leads to triangular-shaped nucleus with a small width and large nucleation barrier.~\cite{Tybell02p97601} The depolarization energy $U_d$ in the MW model depends on $P_s^2/\varepsilon$, where $P_s$ is the bulk polarization as a sharp polarization change from $P_s \rightarrow -P_s$ is assumed. Therefore, the polarization frustration that actually occurrs at DWs should substantially reduce the depolarization energy because of the smaller polarization gradient thus smaller boundary charges and larger dielectric screening.~\cite{Liu16p360} Similar polarization frustration is likely to occur at ferroelectric/metal interface, and may also play a role in resolving the Landauer's paradox of an implausibly large nucleation energy barrier (10$^8$$k_BT$) for single domain switching. 

Distinguishing and understanding the origins of various contributions to ferroelectric susceptibility is critical for controlled materials design and performance optimization. Our first-principles and molecular dynamics simulations demonstrate that the frustrated dipoles within the DWs acquire larger dipole fluctuation and are responsible for the enhanced dielectric response of stationary DWs under subswitching field. Our work suggests that the dielectric constant at the DW can be two times higher than the defect-free bulk. The mechanism applies to {\em all} types of DWs, as polarization suppression at domain boundaries is the natural consequence of reducing gradient energy penalty. Exploring how defect pinning of DW may influence the stationary DW contribution will be a useful future research. 

\newpage
\noindent {\bf Acknowledgements} This work is partly supported by US ONR. 
SL and REC are supported by the Carnegie Institution for Science. REC is also supported by the
ERC Advanced Grant ToMCaT. Computational support was provided by the US DOD through a Challenge Grant from the HPCMO.
%\bibliography{SL}

\begin{thebibliography}{64}%
\makeatletter
\providecommand \@ifxundefined [1]{%
 \@ifx{#1\undefined}
}%
\providecommand \@ifnum [1]{%
 \ifnum #1\expandafter \@firstoftwo
 \else \expandafter \@secondoftwo
 \fi
}%
\providecommand \@ifx [1]{%
 \ifx #1\expandafter \@firstoftwo
 \else \expandafter \@secondoftwo
 \fi
}%
\providecommand \natexlab [1]{#1}%
\providecommand \enquote  [1]{``#1''}%
\providecommand \bibnamefont  [1]{#1}%
\providecommand \bibfnamefont [1]{#1}%
\providecommand \citenamefont [1]{#1}%
\providecommand \href@noop [0]{\@secondoftwo}%
\providecommand \href [0]{\begingroup \@sanitize@url \@href}%
\providecommand \@href[1]{\@@startlink{#1}\@@href}%
\providecommand \@@href[1]{\endgroup#1\@@endlink}%
\providecommand \@sanitize@url [0]{\catcode `\\12\catcode `\$12\catcode
  `\&12\catcode `\#12\catcode `\^12\catcode `\_12\catcode `\%12\relax}%
\providecommand \@@startlink[1]{}%
\providecommand \@@endlink[0]{}%
\providecommand \url  [0]{\begingroup\@sanitize@url \@url }%
\providecommand \@url [1]{\endgroup\@href {#1}{\urlprefix }}%
\providecommand \urlprefix  [0]{URL }%
\providecommand \Eprint [0]{\href }%
\providecommand \doibase [0]{http://dx.doi.org/}%
\providecommand \selectlanguage [0]{\@gobble}%
\providecommand \bibinfo  [0]{\@secondoftwo}%
\providecommand \bibfield  [0]{\@secondoftwo}%
\providecommand \translation [1]{[#1]}%
\providecommand \BibitemOpen [0]{}%
\providecommand \bibitemStop [0]{}%
\providecommand \bibitemNoStop [0]{.\EOS\space}%
\providecommand \EOS [0]{\spacefactor3000\relax}%
\providecommand \BibitemShut  [1]{\csname bibitem#1\endcsname}%
\let\auto@bib@innerbib\@empty
%</preamble>
\bibitem [{\citenamefont {Scott}(2007)}]{Scott07p954}%
  \BibitemOpen
  \bibfield  {author} {\bibinfo {author} {\bibfnamefont {J.~F.}\ \bibnamefont
  {Scott}},\ }\href@noop {} {\bibfield  {journal} {\bibinfo  {journal}
  {Science}\ }\textbf {\bibinfo {volume} {315}},\ \bibinfo {pages} {954}
  (\bibinfo {year} {2007})}\BibitemShut {NoStop}%
\bibitem [{\citenamefont {Pramanick}\ \emph {et~al.}(2012)\citenamefont
  {Pramanick}, \citenamefont {Prewitt}, \citenamefont {Forrester},\ and\
  \citenamefont {Jones}}]{Pramanick12p243}%
  \BibitemOpen
  \bibfield  {author} {\bibinfo {author} {\bibfnamefont {A.}~\bibnamefont
  {Pramanick}}, \bibinfo {author} {\bibfnamefont {A.~D.}\ \bibnamefont
  {Prewitt}}, \bibinfo {author} {\bibfnamefont {J.~S.}\ \bibnamefont
  {Forrester}}, \ and\ \bibinfo {author} {\bibfnamefont {J.~L.}\ \bibnamefont
  {Jones}},\ }\href@noop {} {\bibfield  {journal} {\bibinfo  {journal} {Crit.
  Rev. Solid State Mater. Sci.}\ }\textbf {\bibinfo {volume} {37}},\ \bibinfo
  {pages} {243} (\bibinfo {year} {2012})}\BibitemShut {NoStop}%
\bibitem [{\citenamefont {Zhang}\ \emph {et~al.}(1994)\citenamefont {Zhang},
  \citenamefont {Wang}, \citenamefont {Kim},\ and\ \citenamefont
  {Cross}}]{Zhang94p454}%
  \BibitemOpen
  \bibfield  {author} {\bibinfo {author} {\bibfnamefont {Q.}~\bibnamefont
  {Zhang}}, \bibinfo {author} {\bibfnamefont {H.}~\bibnamefont {Wang}},
  \bibinfo {author} {\bibfnamefont {N.}~\bibnamefont {Kim}}, \ and\ \bibinfo
  {author} {\bibfnamefont {L.}~\bibnamefont {Cross}},\ }\href@noop {}
  {\bibfield  {journal} {\bibinfo  {journal} {J. Appl. Phys.}\ }\textbf
  {\bibinfo {volume} {75}},\ \bibinfo {pages} {454} (\bibinfo {year}
  {1994})}\BibitemShut {NoStop}%
\bibitem [{\citenamefont {Taylor}\ and\ \citenamefont
  {Damjanovic}(1997)}]{Taylor97p1973}%
  \BibitemOpen
  \bibfield  {author} {\bibinfo {author} {\bibfnamefont {D.}~\bibnamefont
  {Taylor}}\ and\ \bibinfo {author} {\bibfnamefont {D.}~\bibnamefont
  {Damjanovic}},\ }\href@noop {} {\bibfield  {journal} {\bibinfo  {journal} {J.
  Appl. Phys.}\ }\textbf {\bibinfo {volume} {82}},\ \bibinfo {pages} {1973}
  (\bibinfo {year} {1997})}\BibitemShut {NoStop}%
\bibitem [{\citenamefont {Xu}\ \emph {et~al.}(2001)\citenamefont {Xu},
  \citenamefont {Trolier-McKinstry}, \citenamefont {Ren}, \citenamefont {Xu},
  \citenamefont {Xie},\ and\ \citenamefont {Hemker}}]{Xu01p1336}%
  \BibitemOpen
  \bibfield  {author} {\bibinfo {author} {\bibfnamefont {F.}~\bibnamefont
  {Xu}}, \bibinfo {author} {\bibfnamefont {S.}~\bibnamefont
  {Trolier-McKinstry}}, \bibinfo {author} {\bibfnamefont {W.}~\bibnamefont
  {Ren}}, \bibinfo {author} {\bibfnamefont {B.}~\bibnamefont {Xu}}, \bibinfo
  {author} {\bibfnamefont {Z.-L.}\ \bibnamefont {Xie}}, \ and\ \bibinfo
  {author} {\bibfnamefont {K.}~\bibnamefont {Hemker}},\ }\href
  {http://dx.doi.org/10.1063/1.1325005} {\bibfield  {journal} {\bibinfo
  {journal} {J. Appl. Phys.}\ }\textbf {\bibinfo {volume} {89}},\ \bibinfo
  {pages} {1336} (\bibinfo {year} {2001})}\BibitemShut {NoStop}%
\bibitem [{\citenamefont {Xu}\ \emph {et~al.}(2014)\citenamefont {Xu},
  \citenamefont {Karthik}, \citenamefont {Damodaran},\ and\ \citenamefont
  {Martin}}]{Xu14p3120}%
  \BibitemOpen
  \bibfield  {author} {\bibinfo {author} {\bibfnamefont {R.}~\bibnamefont
  {Xu}}, \bibinfo {author} {\bibfnamefont {J.}~\bibnamefont {Karthik}},
  \bibinfo {author} {\bibfnamefont {A.~R.}\ \bibnamefont {Damodaran}}, \ and\
  \bibinfo {author} {\bibfnamefont {L.~W.}\ \bibnamefont {Martin}},\
  }\href@noop {} {\bibfield  {journal} {\bibinfo  {journal} {Nat. Comm.}\
  }\textbf {\bibinfo {volume} {5}},\ \bibinfo {pages} {3120} (\bibinfo {year}
  {2014})}\BibitemShut {NoStop}%
\bibitem [{\citenamefont {Karthik}\ and\ \citenamefont
  {Martin}(2011)}]{Karthik11p024102}%
  \BibitemOpen
  \bibfield  {author} {\bibinfo {author} {\bibfnamefont {J.}~\bibnamefont
  {Karthik}}\ and\ \bibinfo {author} {\bibfnamefont {L.}~\bibnamefont
  {Martin}},\ }\href@noop {} {\bibfield  {journal} {\bibinfo  {journal} {Phys.
  Rev. B}\ }\textbf {\bibinfo {volume} {84}},\ \bibinfo {pages} {024102}
  (\bibinfo {year} {2011})}\BibitemShut {NoStop}%
\bibitem [{\citenamefont {Karthik}\ \emph {et~al.}(2012)\citenamefont
  {Karthik}, \citenamefont {Damodaran},\ and\ \citenamefont
  {Martin}}]{Karthik12p167601}%
  \BibitemOpen
  \bibfield  {author} {\bibinfo {author} {\bibfnamefont {J.}~\bibnamefont
  {Karthik}}, \bibinfo {author} {\bibfnamefont {A.}~\bibnamefont {Damodaran}},
  \ and\ \bibinfo {author} {\bibfnamefont {L.}~\bibnamefont {Martin}},\
  }\href@noop {} {\bibfield  {journal} {\bibinfo  {journal} {Phys. Rev. Lett.}\
  }\textbf {\bibinfo {volume} {108}},\ \bibinfo {pages} {167601} (\bibinfo
  {year} {2012})}\BibitemShut {NoStop}%
\bibitem [{\citenamefont {Zubko}\ \emph {et~al.}(2016)\citenamefont {Zubko},
  \citenamefont {Wojde{\l}}, \citenamefont {Hadjimichael}, \citenamefont
  {Fernandez-Pena}, \citenamefont {Sen{\'{e}}}, \citenamefont {Luk'yanchuk},
  \citenamefont {Triscone},\ and\ \citenamefont
  {{\'{I}}{\~{n}}iguez}}]{Zubko16p524}%
  \BibitemOpen
  \bibfield  {author} {\bibinfo {author} {\bibfnamefont {P.}~\bibnamefont
  {Zubko}}, \bibinfo {author} {\bibfnamefont {J.~C.}\ \bibnamefont
  {Wojde{\l}}}, \bibinfo {author} {\bibfnamefont {M.}~\bibnamefont
  {Hadjimichael}}, \bibinfo {author} {\bibfnamefont {S.}~\bibnamefont
  {Fernandez-Pena}}, \bibinfo {author} {\bibfnamefont {A.}~\bibnamefont
  {Sen{\'{e}}}}, \bibinfo {author} {\bibfnamefont {I.}~\bibnamefont
  {Luk'yanchuk}}, \bibinfo {author} {\bibfnamefont {J.-M.}\ \bibnamefont
  {Triscone}}, \ and\ \bibinfo {author} {\bibfnamefont {J.}~\bibnamefont
  {{\'{I}}{\~{n}}iguez}},\ }\href {\doibase 10.1038/nature17659} {\bibfield
  {journal} {\bibinfo  {journal} {Nature}\ }\textbf {\bibinfo {volume} {534}},\
  \bibinfo {pages} {524} (\bibinfo {year} {2016})}\BibitemShut {NoStop}%
\bibitem [{\citenamefont {Pramanick}\ \emph {et~al.}(2011)\citenamefont
  {Pramanick}, \citenamefont {Damjanovic}, \citenamefont {Daniels},
  \citenamefont {Nino},\ and\ \citenamefont {Jones}}]{Pramanick11p293}%
  \BibitemOpen
  \bibfield  {author} {\bibinfo {author} {\bibfnamefont {A.}~\bibnamefont
  {Pramanick}}, \bibinfo {author} {\bibfnamefont {D.}~\bibnamefont
  {Damjanovic}}, \bibinfo {author} {\bibfnamefont {J.~E.}\ \bibnamefont
  {Daniels}}, \bibinfo {author} {\bibfnamefont {J.~C.}\ \bibnamefont {Nino}}, \
  and\ \bibinfo {author} {\bibfnamefont {J.~L.}\ \bibnamefont {Jones}},\ }\href
  {\doibase 10.1111/j.1551-2916.2010.04240.x} {\bibfield  {journal} {\bibinfo
  {journal} {J. Am. Ceram. Soc.}\ }\textbf {\bibinfo {volume} {94}},\ \bibinfo
  {pages} {293} (\bibinfo {year} {2011})}\BibitemShut {NoStop}%
\bibitem [{\citenamefont {Zednik}\ \emph {et~al.}(2011)\citenamefont {Zednik},
  \citenamefont {Varatharajan}, \citenamefont {Oliver}, \citenamefont
  {Valanoor},\ and\ \citenamefont {McIntyre}}]{Zednik11p3104}%
  \BibitemOpen
  \bibfield  {author} {\bibinfo {author} {\bibfnamefont {R.~J.}\ \bibnamefont
  {Zednik}}, \bibinfo {author} {\bibfnamefont {A.}~\bibnamefont
  {Varatharajan}}, \bibinfo {author} {\bibfnamefont {M.}~\bibnamefont
  {Oliver}}, \bibinfo {author} {\bibfnamefont {N.}~\bibnamefont {Valanoor}}, \
  and\ \bibinfo {author} {\bibfnamefont {P.~C.}\ \bibnamefont {McIntyre}},\
  }\href {\doibase 10.1002/adfm.201100445} {\bibfield  {journal} {\bibinfo
  {journal} {Adv. Funct. Mater.}\ }\textbf {\bibinfo {volume} {21}},\ \bibinfo
  {pages} {3104} (\bibinfo {year} {2011})}\BibitemShut {NoStop}%
\bibitem [{\citenamefont {Lawless}\ and\ \citenamefont
  {Fousek}(1970)}]{Lawless70p419}%
  \BibitemOpen
  \bibfield  {author} {\bibinfo {author} {\bibfnamefont {W.~N.}\ \bibnamefont
  {Lawless}}\ and\ \bibinfo {author} {\bibfnamefont {J.}~\bibnamefont
  {Fousek}},\ }\href {\doibase 10.1143/jpsj.28.419} {\bibfield  {journal}
  {\bibinfo  {journal} {J. Phys. Soc. Jpn.}\ }\textbf {\bibinfo {volume}
  {28}},\ \bibinfo {pages} {419} (\bibinfo {year} {1970})}\BibitemShut
  {NoStop}%
\bibitem [{\citenamefont {Marvan}(1969)}]{Marvan69p482}%
  \BibitemOpen
  \bibfield  {author} {\bibinfo {author} {\bibfnamefont {M.}~\bibnamefont
  {Marvan}},\ }\href {\doibase 10.1007/bf01691813} {\bibfield  {journal}
  {\bibinfo  {journal} {Czechoslovak Journal of Physics}\ }\textbf {\bibinfo
  {volume} {19}},\ \bibinfo {pages} {482} (\bibinfo {year} {1969})}\BibitemShut
  {NoStop}%
\bibitem [{\citenamefont {Liu}\ \emph {et~al.}(2016)\citenamefont {Liu},
  \citenamefont {Grinberg},\ and\ \citenamefont {Rappe}}]{Liu16p360}%
  \BibitemOpen
  \bibfield  {author} {\bibinfo {author} {\bibfnamefont {S.}~\bibnamefont
  {Liu}}, \bibinfo {author} {\bibfnamefont {I.}~\bibnamefont {Grinberg}}, \
  and\ \bibinfo {author} {\bibfnamefont {A.~M.}\ \bibnamefont {Rappe}},\ }\href
  {\doibase 10.1038/nature18286} {\bibfield  {journal} {\bibinfo  {journal}
  {Nature}\ }\textbf {\bibinfo {volume} {534}},\ \bibinfo {pages} {360}
  (\bibinfo {year} {2016})}\BibitemShut {NoStop}%
\bibitem [{\citenamefont {Tagantsev}\ \emph {et~al.}(2010)\citenamefont
  {Tagantsev}, \citenamefont {Cross},\ and\ \citenamefont
  {Fousek}}]{Tagantsev10Book}%
  \BibitemOpen
  \bibfield  {author} {\bibinfo {author} {\bibfnamefont {A.~K.}\ \bibnamefont
  {Tagantsev}}, \bibinfo {author} {\bibfnamefont {L.~E.}\ \bibnamefont
  {Cross}}, \ and\ \bibinfo {author} {\bibfnamefont {J.}~\bibnamefont
  {Fousek}},\ }\href@noop {} {\emph {\bibinfo {title} {Domains in Ferroic
  Crystals and Thin Films}}}\ (\bibinfo  {publisher} {Springer},\ \bibinfo
  {year} {2010})\BibitemShut {NoStop}%
\bibitem [{\citenamefont {Kittel}(1951)}]{Kittel51p458}%
  \BibitemOpen
  \bibfield  {author} {\bibinfo {author} {\bibfnamefont {C.}~\bibnamefont
  {Kittel}},\ }\href@noop {} {\bibfield  {journal} {\bibinfo  {journal} {Phys.
  Rev.}\ }\textbf {\bibinfo {volume} {83}},\ \bibinfo {pages} {458} (\bibinfo
  {year} {1951})}\BibitemShut {NoStop}%
\bibitem [{\citenamefont {Fousek}\ and\ \citenamefont
  {Brezina}(1964)}]{Fousek64p830}%
  \BibitemOpen
  \bibfield  {author} {\bibinfo {author} {\bibfnamefont {J.}~\bibnamefont
  {Fousek}}\ and\ \bibinfo {author} {\bibfnamefont {B.}~\bibnamefont
  {Brezina}},\ }\href@noop {} {\bibfield  {journal} {\bibinfo  {journal} {J.
  Phys. Soc. Jpn.}\ }\textbf {\bibinfo {volume} {19}},\ \bibinfo {pages} {830}
  (\bibinfo {year} {1964})}\BibitemShut {NoStop}%
\bibitem [{\citenamefont {Arlt}\ \emph {et~al.}(1987)\citenamefont {Arlt},
  \citenamefont {Dederichs},\ and\ \citenamefont {Herbiet}}]{Arlt87p37}%
  \BibitemOpen
  \bibfield  {author} {\bibinfo {author} {\bibfnamefont {G.}~\bibnamefont
  {Arlt}}, \bibinfo {author} {\bibfnamefont {H.}~\bibnamefont {Dederichs}}, \
  and\ \bibinfo {author} {\bibfnamefont {R.}~\bibnamefont {Herbiet}},\ }\href
  {\doibase 10.1080/00150198708014493} {\bibfield  {journal} {\bibinfo
  {journal} {Ferroelectrics}\ }\textbf {\bibinfo {volume} {74}},\ \bibinfo
  {pages} {37} (\bibinfo {year} {1987})}\BibitemShut {NoStop}%
\bibitem [{\citenamefont {Arlt}\ and\ \citenamefont
  {Pertsev}(1991)}]{Arlt91p2283}%
  \BibitemOpen
  \bibfield  {author} {\bibinfo {author} {\bibfnamefont {G.}~\bibnamefont
  {Arlt}}\ and\ \bibinfo {author} {\bibfnamefont {N.~A.}\ \bibnamefont
  {Pertsev}},\ }\href {\doibase 10.1063/1.349421} {\bibfield  {journal}
  {\bibinfo  {journal} {J. Appl. Phys.}\ }\textbf {\bibinfo {volume} {70}},\
  \bibinfo {pages} {2283} (\bibinfo {year} {1991})}\BibitemShut {NoStop}%
\bibitem [{\citenamefont {Pertsev}\ \emph {et~al.}(1995)\citenamefont
  {Pertsev}, \citenamefont {Arlt},\ and\ \citenamefont
  {Zembilgotov}}]{Pertsev95p135}%
  \BibitemOpen
  \bibfield  {author} {\bibinfo {author} {\bibfnamefont {N.~A.}\ \bibnamefont
  {Pertsev}}, \bibinfo {author} {\bibfnamefont {G.}~\bibnamefont {Arlt}}, \
  and\ \bibinfo {author} {\bibfnamefont {A.~G.}\ \bibnamefont {Zembilgotov}},\
  }\href {\doibase 10.1016/0167-9317(95)00131-X} {\bibfield  {journal}
  {\bibinfo  {journal} {Microelectron. Eng.}\ }\textbf {\bibinfo {volume}
  {29}},\ \bibinfo {pages} {135} (\bibinfo {year} {1995})}\BibitemShut
  {NoStop}%
\bibitem [{\citenamefont {Pertsev}\ \emph {et~al.}(1996)\citenamefont
  {Pertsev}, \citenamefont {Arlt},\ and\ \citenamefont
  {Zembilgotov}}]{Pertsev96p1364}%
  \BibitemOpen
  \bibfield  {author} {\bibinfo {author} {\bibfnamefont {N.~A.}\ \bibnamefont
  {Pertsev}}, \bibinfo {author} {\bibfnamefont {G.}~\bibnamefont {Arlt}}, \
  and\ \bibinfo {author} {\bibfnamefont {A.~G.}\ \bibnamefont {Zembilgotov}},\
  }\href {\doibase 10.1103/physrevlett.76.1364} {\bibfield  {journal} {\bibinfo
   {journal} {Phys. Rev. Lett.}\ }\textbf {\bibinfo {volume} {76}},\ \bibinfo
  {pages} {1364} (\bibinfo {year} {1996})}\BibitemShut {NoStop}%
\bibitem [{\citenamefont {Kim}\ \emph {et~al.}(2010)\citenamefont {Kim},
  \citenamefont {Han}, \citenamefont {Lee}, \citenamefont {Baik}, \citenamefont
  {Hesse},\ and\ \citenamefont {Alexe}}]{Kim10p1266}%
  \BibitemOpen
  \bibfield  {author} {\bibinfo {author} {\bibfnamefont {Y.}~\bibnamefont
  {Kim}}, \bibinfo {author} {\bibfnamefont {H.}~\bibnamefont {Han}}, \bibinfo
  {author} {\bibfnamefont {W.}~\bibnamefont {Lee}}, \bibinfo {author}
  {\bibfnamefont {S.}~\bibnamefont {Baik}}, \bibinfo {author} {\bibfnamefont
  {D.}~\bibnamefont {Hesse}}, \ and\ \bibinfo {author} {\bibfnamefont
  {M.}~\bibnamefont {Alexe}},\ }\href@noop {} {\bibfield  {journal} {\bibinfo
  {journal} {Nano Lett.}\ }\textbf {\bibinfo {volume} {10}},\ \bibinfo {pages}
  {1266} (\bibinfo {year} {2010})}\BibitemShut {NoStop}%
\bibitem [{\citenamefont {Dawber}\ \emph {et~al.}(2003)\citenamefont {Dawber},
  \citenamefont {Jung},\ and\ \citenamefont {Scott}}]{Dawber03p436}%
  \BibitemOpen
  \bibfield  {author} {\bibinfo {author} {\bibfnamefont {M.}~\bibnamefont
  {Dawber}}, \bibinfo {author} {\bibfnamefont {D.~J.}\ \bibnamefont {Jung}}, \
  and\ \bibinfo {author} {\bibfnamefont {J.~F.}\ \bibnamefont {Scott}},\
  }\href@noop {} {\bibfield  {journal} {\bibinfo  {journal} {Appl. Phys.
  Lett.}\ }\textbf {\bibinfo {volume} {82}},\ \bibinfo {pages} {436} (\bibinfo
  {year} {2003})}\BibitemShut {NoStop}%
\bibitem [{\citenamefont {Sharma}\ \emph {et~al.}(2013)\citenamefont {Sharma},
  \citenamefont {McQuaid}, \citenamefont {McGilly}, \citenamefont {Gregg},\
  and\ \citenamefont {Gruverman}}]{Sharma13p1323}%
  \BibitemOpen
  \bibfield  {author} {\bibinfo {author} {\bibfnamefont {P.}~\bibnamefont
  {Sharma}}, \bibinfo {author} {\bibfnamefont {R.~G.~P.}\ \bibnamefont
  {McQuaid}}, \bibinfo {author} {\bibfnamefont {L.~J.}\ \bibnamefont
  {McGilly}}, \bibinfo {author} {\bibfnamefont {J.~M.}\ \bibnamefont {Gregg}},
  \ and\ \bibinfo {author} {\bibfnamefont {A.}~\bibnamefont {Gruverman}},\
  }\href@noop {} {\bibfield  {journal} {\bibinfo  {journal} {Adv. Mater.}\
  }\textbf {\bibinfo {volume} {25}},\ \bibinfo {pages} {1323} (\bibinfo {year}
  {2013})}\BibitemShut {NoStop}%
\bibitem [{\citenamefont {Molotskii}\ \emph {et~al.}(2007)\citenamefont
  {Molotskii}, \citenamefont {Rosenwaks},\ and\ \citenamefont
  {Rosenman}}]{Molotskii07p271}%
  \BibitemOpen
  \bibfield  {author} {\bibinfo {author} {\bibfnamefont {M.}~\bibnamefont
  {Molotskii}}, \bibinfo {author} {\bibfnamefont {Y.}~\bibnamefont
  {Rosenwaks}}, \ and\ \bibinfo {author} {\bibfnamefont {G.}~\bibnamefont
  {Rosenman}},\ }\href@noop {} {\bibfield  {journal} {\bibinfo  {journal}
  {Annu. Rev. Mater. Res.}\ }\textbf {\bibinfo {volume} {37}},\ \bibinfo
  {pages} {271} (\bibinfo {year} {2007})}\BibitemShut {NoStop}%
\bibitem [{\citenamefont {Liu}\ \emph {et~al.}(2013{\natexlab{a}})\citenamefont
  {Liu}, \citenamefont {Grinberg},\ and\ \citenamefont {Rappe}}]{Liu13p232907}%
  \BibitemOpen
  \bibfield  {author} {\bibinfo {author} {\bibfnamefont {S.}~\bibnamefont
  {Liu}}, \bibinfo {author} {\bibfnamefont {I.}~\bibnamefont {Grinberg}}, \
  and\ \bibinfo {author} {\bibfnamefont {A.~M.}\ \bibnamefont {Rappe}},\
  }\href@noop {} {\bibfield  {journal} {\bibinfo  {journal} {Appl. Phys.
  Lett.}\ }\textbf {\bibinfo {volume} {103}},\ \bibinfo {pages} {232907}
  (\bibinfo {year} {2013}{\natexlab{a}})}\BibitemShut {NoStop}%
\bibitem [{\citenamefont {Meyer}\ and\ \citenamefont
  {Vanderbilt}(2002)}]{Meyer02p104111}%
  \BibitemOpen
  \bibfield  {author} {\bibinfo {author} {\bibfnamefont {B.}~\bibnamefont
  {Meyer}}\ and\ \bibinfo {author} {\bibfnamefont {D.}~\bibnamefont
  {Vanderbilt}},\ }\href@noop {} {\bibfield  {journal} {\bibinfo  {journal}
  {Phys. Rev. B}\ }\textbf {\bibinfo {volume} {65}},\ \bibinfo {pages} {104111}
  (\bibinfo {year} {2002})}\BibitemShut {NoStop}%
\bibitem [{\citenamefont {Pöykkö}\ and\ \citenamefont
  {Chadi}(1999)}]{Poykko99p2830}%
  \BibitemOpen
  \bibfield  {author} {\bibinfo {author} {\bibfnamefont {S.}~\bibnamefont
  {Pöykkö}}\ and\ \bibinfo {author} {\bibfnamefont {D.~J.}\ \bibnamefont
  {Chadi}},\ }\href {\doibase 10.1063/1.125164} {\bibfield  {journal} {\bibinfo
   {journal} {Appl. Phys. Lett.}\ }\textbf {\bibinfo {volume} {75}},\ \bibinfo
  {pages} {2830} (\bibinfo {year} {1999})}\BibitemShut {NoStop}%
\bibitem [{\citenamefont {Wojde{\l}}\ and\ \citenamefont
  {{\'{I}}{\~{n}}iguez}(2014)}]{Wojde14p247603}%
  \BibitemOpen
  \bibfield  {author} {\bibinfo {author} {\bibfnamefont {J.~C.}\ \bibnamefont
  {Wojde{\l}}}\ and\ \bibinfo {author} {\bibfnamefont {J.}~\bibnamefont
  {{\'{I}}{\~{n}}iguez}},\ }\href {\doibase 10.1103/physrevlett.112.247603}
  {\bibfield  {journal} {\bibinfo  {journal} {Phys. Rev. Lett.}\ }\textbf
  {\bibinfo {volume} {112}},\ \bibinfo {pages} {247603} (\bibinfo {year}
  {2014})}\BibitemShut {NoStop}%
\bibitem [{\citenamefont {Rabe}\ \emph {et~al.}(2007)\citenamefont {Rabe},
  \citenamefont {Ahn},\ and\ \citenamefont {Triscone}}]{Rabe07Book}%
  \BibitemOpen
  \bibfield  {author} {\bibinfo {author} {\bibfnamefont {K.~M.}\ \bibnamefont
  {Rabe}}, \bibinfo {author} {\bibfnamefont {C.~H.}\ \bibnamefont {Ahn}}, \
  and\ \bibinfo {author} {\bibfnamefont {J.-M.}\ \bibnamefont {Triscone}},\
  }\href@noop {} {\emph {\bibinfo {title} {Physics of ferroelectrics: a modern
  perspective}}},\ Vol.\ \bibinfo {volume} {105}\ (\bibinfo  {publisher}
  {Springer Science \& Business Media},\ \bibinfo {year} {2007})\BibitemShut
  {NoStop}%
\bibitem [{\citenamefont {Bilc}\ \emph {et~al.}(2008)\citenamefont {Bilc},
  \citenamefont {Orlando}, \citenamefont {Shaltaf}, \citenamefont {Rignanese},
  \citenamefont {{\'{I}}{\~{n}}iguez},\ and\ \citenamefont
  {Ghosez}}]{Bilc08p165107}%
  \BibitemOpen
  \bibfield  {author} {\bibinfo {author} {\bibfnamefont {D.~I.}\ \bibnamefont
  {Bilc}}, \bibinfo {author} {\bibfnamefont {R.}~\bibnamefont {Orlando}},
  \bibinfo {author} {\bibfnamefont {R.}~\bibnamefont {Shaltaf}}, \bibinfo
  {author} {\bibfnamefont {G.-M.}\ \bibnamefont {Rignanese}}, \bibinfo {author}
  {\bibfnamefont {J.}~\bibnamefont {{\'{I}}{\~{n}}iguez}}, \ and\ \bibinfo
  {author} {\bibfnamefont {P.}~\bibnamefont {Ghosez}},\ }\href {\doibase
  10.1103/physrevb.77.165107} {\bibfield  {journal} {\bibinfo  {journal} {Phys.
  Rev. B}\ }\textbf {\bibinfo {volume} {77}},\ \bibinfo {pages} {165107}
  (\bibinfo {year} {2008})}\BibitemShut {NoStop}%
\bibitem [{\citenamefont {Sharma}\ \emph {et~al.}(2014)\citenamefont {Sharma},
  \citenamefont {Kreisel},\ and\ \citenamefont {Ghosez}}]{Sharma14p214102}%
  \BibitemOpen
  \bibfield  {author} {\bibinfo {author} {\bibfnamefont {H.}~\bibnamefont
  {Sharma}}, \bibinfo {author} {\bibfnamefont {J.}~\bibnamefont {Kreisel}}, \
  and\ \bibinfo {author} {\bibfnamefont {P.}~\bibnamefont {Ghosez}},\ }\href
  {\doibase 10.1103/physrevb.90.214102} {\bibfield  {journal} {\bibinfo
  {journal} {Phys. Rev. B}\ }\textbf {\bibinfo {volume} {90}},\ \bibinfo
  {pages} {214102} (\bibinfo {year} {2014})}\BibitemShut {NoStop}%
\bibitem [{\citenamefont {Wu}\ and\ \citenamefont {Cohen}(2006)}]{Wu06p235116}%
  \BibitemOpen
  \bibfield  {author} {\bibinfo {author} {\bibfnamefont {Z.}~\bibnamefont
  {Wu}}\ and\ \bibinfo {author} {\bibfnamefont {R.~E.}\ \bibnamefont {Cohen}},\
  }\href {\doibase 10.1103/physrevb.73.235116} {\bibfield  {journal} {\bibinfo
  {journal} {Phys. Rev. B}\ }\textbf {\bibinfo {volume} {73}},\ \bibinfo
  {pages} {235116} (\bibinfo {year} {2006})}\BibitemShut {NoStop}%
\bibitem [{\citenamefont {Zhao}\ and\ \citenamefont
  {Truhlar}(2008)}]{Zhao08p184109}%
  \BibitemOpen
  \bibfield  {author} {\bibinfo {author} {\bibfnamefont {Y.}~\bibnamefont
  {Zhao}}\ and\ \bibinfo {author} {\bibfnamefont {D.~G.}\ \bibnamefont
  {Truhlar}},\ }\href@noop {} {\bibfield  {journal} {\bibinfo  {journal} {J.
  Chem. Phys.}\ }\textbf {\bibinfo {volume} {128}},\ \bibinfo {pages} {184109}
  (\bibinfo {year} {2008})}\BibitemShut {NoStop}%
\bibitem [{\citenamefont {Garrity}\ \emph {et~al.}(2014)\citenamefont
  {Garrity}, \citenamefont {Bennett}, \citenamefont {Rabe},\ and\ \citenamefont
  {Vanderbilt}}]{Garrity14p446}%
  \BibitemOpen
  \bibfield  {author} {\bibinfo {author} {\bibfnamefont {K.~F.}\ \bibnamefont
  {Garrity}}, \bibinfo {author} {\bibfnamefont {J.~W.}\ \bibnamefont
  {Bennett}}, \bibinfo {author} {\bibfnamefont {K.~M.}\ \bibnamefont {Rabe}}, \
  and\ \bibinfo {author} {\bibfnamefont {D.}~\bibnamefont {Vanderbilt}},\
  }\href {\doibase 10.1016/j.commatsci.2013.08.053} {\bibfield  {journal}
  {\bibinfo  {journal} {Comput. Mater. Sci.}\ }\textbf {\bibinfo {volume}
  {81}},\ \bibinfo {pages} {446} (\bibinfo {year} {2014})}\BibitemShut
  {NoStop}%
\bibitem [{\citenamefont {Corso}(2014)}]{DalCorso14p337}%
  \BibitemOpen
  \bibfield  {author} {\bibinfo {author} {\bibfnamefont {A.~D.}\ \bibnamefont
  {Corso}},\ }\href {\doibase 10.1016/j.commatsci.2014.07.043} {\bibfield
  {journal} {\bibinfo  {journal} {Comput. Mater. Sci.}\ }\textbf {\bibinfo
  {volume} {95}},\ \bibinfo {pages} {337} (\bibinfo {year} {2014})}\BibitemShut
  {NoStop}%
\bibitem [{\citenamefont {Rappe}\ \emph {et~al.}(1990)\citenamefont {Rappe},
  \citenamefont {Rabe}, \citenamefont {Kaxiras},\ and\ \citenamefont
  {Joannopoulos}}]{Rappe90p1227}%
  \BibitemOpen
  \bibfield  {author} {\bibinfo {author} {\bibfnamefont {A.~M.}\ \bibnamefont
  {Rappe}}, \bibinfo {author} {\bibfnamefont {K.~M.}\ \bibnamefont {Rabe}},
  \bibinfo {author} {\bibfnamefont {E.}~\bibnamefont {Kaxiras}}, \ and\
  \bibinfo {author} {\bibfnamefont {J.~D.}\ \bibnamefont {Joannopoulos}},\
  }\href@noop {} {\bibfield  {journal} {\bibinfo  {journal} {Phys. Rev. B Rapid
  Comm.}\ }\textbf {\bibinfo {volume} {41}},\ \bibinfo {pages} {1227} (\bibinfo
  {year} {1990})}\BibitemShut {NoStop}%
\bibitem [{\citenamefont {Giannozzi}\ \emph {et~al.}(2009)\citenamefont
  {Giannozzi}, \citenamefont {Baroni}, \citenamefont {Bonini}, \citenamefont
  {Calandra}, \citenamefont {Car}, \citenamefont {Cavazzoni}, \citenamefont
  {Ceresoli}, \citenamefont {Chiarotti}, \citenamefont {Cococcioni},
  \citenamefont {Dabo}, \citenamefont {Corso}, \citenamefont {de~Gironcoli},
  \citenamefont {Fabris}, \citenamefont {Fratesi}, \citenamefont {Gebauer},
  \citenamefont {Gerstmann}, \citenamefont {Gougoussis}, \citenamefont
  {Kokalj}, \citenamefont {Lazzeri}, \citenamefont {Martin-Samos},
  \citenamefont {Marzari}, \citenamefont {Mauri}, \citenamefont {Mazzarello},
  \citenamefont {Paolini}, \citenamefont {Pasquarello}, \citenamefont
  {Paulatto}, \citenamefont {Sbraccia}, \citenamefont {Scandolo}, \citenamefont
  {Sclauzero}, \citenamefont {Seitsonen}, \citenamefont {Smogunov},
  \citenamefont {Umari},\ and\ \citenamefont
  {Wentzcovitch}}]{Giannozzi09p395502}%
  \BibitemOpen
  \bibfield  {author} {\bibinfo {author} {\bibfnamefont {P.}~\bibnamefont
  {Giannozzi}}, \bibinfo {author} {\bibfnamefont {S.}~\bibnamefont {Baroni}},
  \bibinfo {author} {\bibfnamefont {N.}~\bibnamefont {Bonini}}, \bibinfo
  {author} {\bibfnamefont {M.}~\bibnamefont {Calandra}}, \bibinfo {author}
  {\bibfnamefont {R.}~\bibnamefont {Car}}, \bibinfo {author} {\bibfnamefont
  {C.}~\bibnamefont {Cavazzoni}}, \bibinfo {author} {\bibfnamefont
  {D.}~\bibnamefont {Ceresoli}}, \bibinfo {author} {\bibfnamefont {G.~L.}\
  \bibnamefont {Chiarotti}}, \bibinfo {author} {\bibfnamefont {M.}~\bibnamefont
  {Cococcioni}}, \bibinfo {author} {\bibfnamefont {I.}~\bibnamefont {Dabo}},
  \bibinfo {author} {\bibfnamefont {A.~D.}\ \bibnamefont {Corso}}, \bibinfo
  {author} {\bibfnamefont {S.}~\bibnamefont {de~Gironcoli}}, \bibinfo {author}
  {\bibfnamefont {S.}~\bibnamefont {Fabris}}, \bibinfo {author} {\bibfnamefont
  {G.}~\bibnamefont {Fratesi}}, \bibinfo {author} {\bibfnamefont
  {R.}~\bibnamefont {Gebauer}}, \bibinfo {author} {\bibfnamefont
  {U.}~\bibnamefont {Gerstmann}}, \bibinfo {author} {\bibfnamefont
  {C.}~\bibnamefont {Gougoussis}}, \bibinfo {author} {\bibfnamefont
  {A.}~\bibnamefont {Kokalj}}, \bibinfo {author} {\bibfnamefont
  {M.}~\bibnamefont {Lazzeri}}, \bibinfo {author} {\bibfnamefont
  {L.}~\bibnamefont {Martin-Samos}}, \bibinfo {author} {\bibfnamefont
  {N.}~\bibnamefont {Marzari}}, \bibinfo {author} {\bibfnamefont
  {F.}~\bibnamefont {Mauri}}, \bibinfo {author} {\bibfnamefont
  {R.}~\bibnamefont {Mazzarello}}, \bibinfo {author} {\bibfnamefont
  {S.}~\bibnamefont {Paolini}}, \bibinfo {author} {\bibfnamefont
  {A.}~\bibnamefont {Pasquarello}}, \bibinfo {author} {\bibfnamefont
  {L.}~\bibnamefont {Paulatto}}, \bibinfo {author} {\bibfnamefont
  {C.}~\bibnamefont {Sbraccia}}, \bibinfo {author} {\bibfnamefont
  {S.}~\bibnamefont {Scandolo}}, \bibinfo {author} {\bibfnamefont
  {G.}~\bibnamefont {Sclauzero}}, \bibinfo {author} {\bibfnamefont {A.~P.}\
  \bibnamefont {Seitsonen}}, \bibinfo {author} {\bibfnamefont {A.}~\bibnamefont
  {Smogunov}}, \bibinfo {author} {\bibfnamefont {P.}~\bibnamefont {Umari}}, \
  and\ \bibinfo {author} {\bibfnamefont {R.~M.}\ \bibnamefont {Wentzcovitch}},\
  }\href@noop {} {\bibfield  {journal} {\bibinfo  {journal} {J. Phys.: Condens.
  Matter}\ }\textbf {\bibinfo {volume} {21}},\ \bibinfo {pages} {395502}
  (\bibinfo {year} {2009})}\BibitemShut {NoStop}%
\bibitem [{\citenamefont {Gonze}\ \emph {et~al.}(2009)\citenamefont {Gonze},
  \citenamefont {Amadon}, \citenamefont {Anglade}, \citenamefont {Beuken},
  \citenamefont {Bottin}, \citenamefont {Boulanger}, \citenamefont {Bruneval},
  \citenamefont {Caliste}, \citenamefont {Caracas}, \citenamefont {Cote},
  \citenamefont {Deutsch}, \citenamefont {Genovese}, \citenamefont {Ghosez},
  \citenamefont {Giantomassi}, \citenamefont {Goedecker}, \citenamefont
  {Hamann}, \citenamefont {Hermet}, \citenamefont {Jollet}, \citenamefont
  {Jomard}, \citenamefont {Leroux}, \citenamefont {Mancini}, \citenamefont
  {Mazevet}, \citenamefont {Oliveira}, \citenamefont {Onida}, \citenamefont
  {Pouillon}, \citenamefont {Rangel}, \citenamefont {Rignanese}, \citenamefont
  {Sangalli}, \citenamefont {Shaltaf}, \citenamefont {Torrent}, \citenamefont
  {Verstraete}, \citenamefont {Zerah}, \citenamefont {Gonze}, \citenamefont
  {Amadon}, \citenamefont {Anglade}, \citenamefont {Beuken}, \citenamefont
  {Bottin}, \citenamefont {Boulanger}, \citenamefont {and' D.~Caliste},
  \citenamefont {Caracas}, \citenamefont {Cote}, \citenamefont {Deutsch},
  \citenamefont {Genovese}, \citenamefont {Ghosez}, \citenamefont
  {Giantomassi}, \citenamefont {Goedecker}, \citenamefont {Hamann},
  \citenamefont {Hermet}, \citenamefont {Jollet}, \citenamefont {Jomard},
  \citenamefont {Leroux}, \citenamefont {Mancini}, \citenamefont {Mazevet},
  \citenamefont {Oliveira}, \citenamefont {Onida}, \citenamefont {Pouillon},
  \citenamefont {Rangel}, \citenamefont {Rignanese}, \citenamefont {Sangalli},
  \citenamefont {Shaltaf}, \citenamefont {Torrent}, \citenamefont {Verstraete},
  \citenamefont {Zerah}, \citenamefont {Gonze}, \citenamefont {Amadon},
  \citenamefont {Anglade}, \citenamefont {Beuken}, \citenamefont {Bottin},
  \citenamefont {Boulanger}, \citenamefont {and' D.~Caliste}, \citenamefont
  {Caracas}, \citenamefont {Cote}, \citenamefont {Deutsch}, \citenamefont
  {Genovese}, \citenamefont {Ghosez}, \citenamefont {Giantomassi},
  \citenamefont {Goedecker}, \citenamefont {Hamann}, \citenamefont {Hermet},
  \citenamefont {Jollet}, \citenamefont {Jomard}, \citenamefont {Leroux},
  \citenamefont {Mancini}, \citenamefont {Mazevet}, \citenamefont {Oliveira},
  \citenamefont {Onida}, \citenamefont {Pouillon}, \citenamefont {Rangel},
  \citenamefont {Rignanese}, \citenamefont {Sangalli}, \citenamefont {Shaltaf},
  \citenamefont {Torrent}, \citenamefont {Verstraete}, \citenamefont {Zerah},\
  and\ \citenamefont {Zwanziger}}]{Gonze09p2582}%
  \BibitemOpen
  \bibfield  {author} {\bibinfo {author} {\bibfnamefont {X.}~\bibnamefont
  {Gonze}}, \bibinfo {author} {\bibfnamefont {B.}~\bibnamefont {Amadon}},
  \bibinfo {author} {\bibfnamefont {P.-M.}\ \bibnamefont {Anglade}}, \bibinfo
  {author} {\bibfnamefont {J.-M.}\ \bibnamefont {Beuken}}, \bibinfo {author}
  {\bibfnamefont {F.}~\bibnamefont {Bottin}}, \bibinfo {author} {\bibfnamefont
  {P.}~\bibnamefont {Boulanger}}, \bibinfo {author} {\bibfnamefont
  {F.}~\bibnamefont {Bruneval}}, \bibinfo {author} {\bibfnamefont
  {D.}~\bibnamefont {Caliste}}, \bibinfo {author} {\bibfnamefont
  {R.}~\bibnamefont {Caracas}}, \bibinfo {author} {\bibfnamefont
  {M.}~\bibnamefont {Cote}}, \bibinfo {author} {\bibfnamefont {T.}~\bibnamefont
  {Deutsch}}, \bibinfo {author} {\bibfnamefont {L.}~\bibnamefont {Genovese}},
  \bibinfo {author} {\bibfnamefont {P.}~\bibnamefont {Ghosez}}, \bibinfo
  {author} {\bibfnamefont {M.}~\bibnamefont {Giantomassi}}, \bibinfo {author}
  {\bibfnamefont {S.}~\bibnamefont {Goedecker}}, \bibinfo {author}
  {\bibfnamefont {D.}~\bibnamefont {Hamann}}, \bibinfo {author} {\bibfnamefont
  {P.}~\bibnamefont {Hermet}}, \bibinfo {author} {\bibfnamefont
  {F.}~\bibnamefont {Jollet}}, \bibinfo {author} {\bibfnamefont
  {G.}~\bibnamefont {Jomard}}, \bibinfo {author} {\bibfnamefont
  {S.}~\bibnamefont {Leroux}}, \bibinfo {author} {\bibfnamefont
  {M.}~\bibnamefont {Mancini}}, \bibinfo {author} {\bibfnamefont
  {S.}~\bibnamefont {Mazevet}}, \bibinfo {author} {\bibfnamefont
  {M.}~\bibnamefont {Oliveira}}, \bibinfo {author} {\bibfnamefont
  {G.}~\bibnamefont {Onida}}, \bibinfo {author} {\bibfnamefont
  {Y.}~\bibnamefont {Pouillon}}, \bibinfo {author} {\bibfnamefont
  {T.}~\bibnamefont {Rangel}}, \bibinfo {author} {\bibfnamefont {G.-M.}\
  \bibnamefont {Rignanese}}, \bibinfo {author} {\bibfnamefont {D.}~\bibnamefont
  {Sangalli}}, \bibinfo {author} {\bibfnamefont {R.}~\bibnamefont {Shaltaf}},
  \bibinfo {author} {\bibfnamefont {M.}~\bibnamefont {Torrent}}, \bibinfo
  {author} {\bibfnamefont {M.}~\bibnamefont {Verstraete}}, \bibinfo {author}
  {\bibfnamefont {G.}~\bibnamefont {Zerah}}, \bibinfo {author} {\bibfnamefont
  {J.~Z.~X.}\ \bibnamefont {Gonze}}, \bibinfo {author} {\bibfnamefont
  {B.}~\bibnamefont {Amadon}}, \bibinfo {author} {\bibfnamefont {P.-M.}\
  \bibnamefont {Anglade}}, \bibinfo {author} {\bibfnamefont {J.-M.}\
  \bibnamefont {Beuken}}, \bibinfo {author} {\bibfnamefont {F.}~\bibnamefont
  {Bottin}}, \bibinfo {author} {\bibfnamefont {P.}~\bibnamefont {Boulanger}},
  \bibinfo {author} {\bibfnamefont {F.~B.}\ \bibnamefont {and' D.~Caliste}},
  \bibinfo {author} {\bibfnamefont {R.}~\bibnamefont {Caracas}}, \bibinfo
  {author} {\bibfnamefont {M.}~\bibnamefont {Cote}}, \bibinfo {author}
  {\bibfnamefont {T.}~\bibnamefont {Deutsch}}, \bibinfo {author} {\bibfnamefont
  {L.}~\bibnamefont {Genovese}}, \bibinfo {author} {\bibfnamefont
  {P.}~\bibnamefont {Ghosez}}, \bibinfo {author} {\bibfnamefont
  {M.}~\bibnamefont {Giantomassi}}, \bibinfo {author} {\bibfnamefont
  {S.}~\bibnamefont {Goedecker}}, \bibinfo {author} {\bibfnamefont
  {D.}~\bibnamefont {Hamann}}, \bibinfo {author} {\bibfnamefont
  {P.}~\bibnamefont {Hermet}}, \bibinfo {author} {\bibfnamefont
  {F.}~\bibnamefont {Jollet}}, \bibinfo {author} {\bibfnamefont
  {G.}~\bibnamefont {Jomard}}, \bibinfo {author} {\bibfnamefont
  {S.}~\bibnamefont {Leroux}}, \bibinfo {author} {\bibfnamefont
  {M.}~\bibnamefont {Mancini}}, \bibinfo {author} {\bibfnamefont
  {S.}~\bibnamefont {Mazevet}}, \bibinfo {author} {\bibfnamefont
  {M.}~\bibnamefont {Oliveira}}, \bibinfo {author} {\bibfnamefont
  {G.}~\bibnamefont {Onida}}, \bibinfo {author} {\bibfnamefont
  {Y.}~\bibnamefont {Pouillon}}, \bibinfo {author} {\bibfnamefont
  {T.}~\bibnamefont {Rangel}}, \bibinfo {author} {\bibfnamefont {G.-M.}\
  \bibnamefont {Rignanese}}, \bibinfo {author} {\bibfnamefont {D.}~\bibnamefont
  {Sangalli}}, \bibinfo {author} {\bibfnamefont {R.}~\bibnamefont {Shaltaf}},
  \bibinfo {author} {\bibfnamefont {M.}~\bibnamefont {Torrent}}, \bibinfo
  {author} {\bibfnamefont {M.}~\bibnamefont {Verstraete}}, \bibinfo {author}
  {\bibfnamefont {G.}~\bibnamefont {Zerah}}, \bibinfo {author} {\bibfnamefont
  {J.~Z.~X.}\ \bibnamefont {Gonze}}, \bibinfo {author} {\bibfnamefont
  {B.}~\bibnamefont {Amadon}}, \bibinfo {author} {\bibfnamefont {P.-M.}\
  \bibnamefont {Anglade}}, \bibinfo {author} {\bibfnamefont {J.-M.}\
  \bibnamefont {Beuken}}, \bibinfo {author} {\bibfnamefont {F.}~\bibnamefont
  {Bottin}}, \bibinfo {author} {\bibfnamefont {P.}~\bibnamefont {Boulanger}},
  \bibinfo {author} {\bibfnamefont {F.~B.}\ \bibnamefont {and' D.~Caliste}},
  \bibinfo {author} {\bibfnamefont {R.}~\bibnamefont {Caracas}}, \bibinfo
  {author} {\bibfnamefont {M.}~\bibnamefont {Cote}}, \bibinfo {author}
  {\bibfnamefont {T.}~\bibnamefont {Deutsch}}, \bibinfo {author} {\bibfnamefont
  {L.}~\bibnamefont {Genovese}}, \bibinfo {author} {\bibfnamefont
  {P.}~\bibnamefont {Ghosez}}, \bibinfo {author} {\bibfnamefont
  {M.}~\bibnamefont {Giantomassi}}, \bibinfo {author} {\bibfnamefont
  {S.}~\bibnamefont {Goedecker}}, \bibinfo {author} {\bibfnamefont
  {D.}~\bibnamefont {Hamann}}, \bibinfo {author} {\bibfnamefont
  {P.}~\bibnamefont {Hermet}}, \bibinfo {author} {\bibfnamefont
  {F.}~\bibnamefont {Jollet}}, \bibinfo {author} {\bibfnamefont
  {G.}~\bibnamefont {Jomard}}, \bibinfo {author} {\bibfnamefont
  {S.}~\bibnamefont {Leroux}}, \bibinfo {author} {\bibfnamefont
  {M.}~\bibnamefont {Mancini}}, \bibinfo {author} {\bibfnamefont
  {S.}~\bibnamefont {Mazevet}}, \bibinfo {author} {\bibfnamefont
  {M.}~\bibnamefont {Oliveira}}, \bibinfo {author} {\bibfnamefont
  {G.}~\bibnamefont {Onida}}, \bibinfo {author} {\bibfnamefont
  {Y.}~\bibnamefont {Pouillon}}, \bibinfo {author} {\bibfnamefont
  {T.}~\bibnamefont {Rangel}}, \bibinfo {author} {\bibfnamefont {G.-M.}\
  \bibnamefont {Rignanese}}, \bibinfo {author} {\bibfnamefont {D.}~\bibnamefont
  {Sangalli}}, \bibinfo {author} {\bibfnamefont {R.}~\bibnamefont {Shaltaf}},
  \bibinfo {author} {\bibfnamefont {M.}~\bibnamefont {Torrent}}, \bibinfo
  {author} {\bibfnamefont {M.}~\bibnamefont {Verstraete}}, \bibinfo {author}
  {\bibfnamefont {G.}~\bibnamefont {Zerah}}, \ and\ \bibinfo {author}
  {\bibfnamefont {J.}~\bibnamefont {Zwanziger}},\ }\href@noop {} {\bibfield
  {journal} {\bibinfo  {journal} {Computer Phys. Commun.}\ }\textbf {\bibinfo
  {volume} {180}},\ \bibinfo {pages} {2582} (\bibinfo {year}
  {2009})}\BibitemShut {NoStop}%
\bibitem [{\citenamefont {Sepliarsky}\ and\ \citenamefont
  {Cohen}(2002)}]{Sepliarsky02p36}%
  \BibitemOpen
  \bibfield  {author} {\bibinfo {author} {\bibfnamefont {M.}~\bibnamefont
  {Sepliarsky}}\ and\ \bibinfo {author} {\bibfnamefont {R.~E.}\ \bibnamefont
  {Cohen}},\ }\href {http://dx.doi.org/10.1063/1.1499550} {\bibfield  {journal}
  {\bibinfo  {journal} {AIP Conf. Proc.}\ }\textbf {\bibinfo {volume} {626}},\
  \bibinfo {pages} {36} (\bibinfo {year} {2002})}\BibitemShut {NoStop}%
\bibitem [{\citenamefont {Sepliarsky}\ and\ \citenamefont
  {Cohen}(2011)}]{Sepliarsky11p435902}%
  \BibitemOpen
  \bibfield  {author} {\bibinfo {author} {\bibfnamefont {M.}~\bibnamefont
  {Sepliarsky}}\ and\ \bibinfo {author} {\bibfnamefont {R.~E.}\ \bibnamefont
  {Cohen}},\ }\href {\doibase 10.1088/0953-8984/23/43/435902} {\bibfield
  {journal} {\bibinfo  {journal} {Journal of Physics: Condensed Matter}\
  }\textbf {\bibinfo {volume} {23}},\ \bibinfo {pages} {435902} (\bibinfo
  {year} {2011})}\BibitemShut {NoStop}%
\bibitem [{\citenamefont {Liu}\ \emph {et~al.}(2013{\natexlab{b}})\citenamefont
  {Liu}, \citenamefont {Grinberg},\ and\ \citenamefont {Rappe}}]{Liu13p102202}%
  \BibitemOpen
  \bibfield  {author} {\bibinfo {author} {\bibfnamefont {S.}~\bibnamefont
  {Liu}}, \bibinfo {author} {\bibfnamefont {I.}~\bibnamefont {Grinberg}}, \
  and\ \bibinfo {author} {\bibfnamefont {A.~M.}\ \bibnamefont {Rappe}},\
  }\href@noop {} {\bibfield  {journal} {\bibinfo  {journal} {J. Physics.:
  Condens. Matter}\ }\textbf {\bibinfo {volume} {25}},\ \bibinfo {pages}
  {102202} (\bibinfo {year} {2013}{\natexlab{b}})}\BibitemShut {NoStop}%
\bibitem [{\citenamefont {Rose}\ and\ \citenamefont
  {Cohen}(2012)}]{Rose12p187604}%
  \BibitemOpen
  \bibfield  {author} {\bibinfo {author} {\bibfnamefont {M.~C.}\ \bibnamefont
  {Rose}}\ and\ \bibinfo {author} {\bibfnamefont {R.~E.}\ \bibnamefont
  {Cohen}},\ }\href@noop {} {\bibfield  {journal} {\bibinfo  {journal} {Phys.
  Rev. Lett.}\ }\textbf {\bibinfo {volume} {109}},\ \bibinfo {pages} {187604}
  (\bibinfo {year} {2012})}\BibitemShut {NoStop}%
\bibitem [{\citenamefont {Zeng}\ and\ \citenamefont
  {Cohen}(2011)}]{Zeng11p142902}%
  \BibitemOpen
  \bibfield  {author} {\bibinfo {author} {\bibfnamefont {X.}~\bibnamefont
  {Zeng}}\ and\ \bibinfo {author} {\bibfnamefont {R.~E.}\ \bibnamefont
  {Cohen}},\ }\href {\doibase 10.1063/1.3646377} {\bibfield  {journal}
  {\bibinfo  {journal} {Appl. Phys. Lett.}\ }\textbf {\bibinfo {volume} {99}},\
  \bibinfo {pages} {142902} (\bibinfo {year} {2011})}\BibitemShut {NoStop}%
\bibitem [{\citenamefont {Hashimoto}\ and\ \citenamefont
  {Moriwake}(2014)}]{Hashimoto14p1074}%
  \BibitemOpen
  \bibfield  {author} {\bibinfo {author} {\bibfnamefont {T.}~\bibnamefont
  {Hashimoto}}\ and\ \bibinfo {author} {\bibfnamefont {H.}~\bibnamefont
  {Moriwake}},\ }\href {\doibase 10.1080/08927022.2014.938067} {\bibfield
  {journal} {\bibinfo  {journal} {Molecular Simulation}\ }\textbf {\bibinfo
  {volume} {41}},\ \bibinfo {pages} {1074} (\bibinfo {year}
  {2014})}\BibitemShut {NoStop}%
\bibitem [{\citenamefont {Sepliarsky}\ \emph {et~al.}(2005)\citenamefont
  {Sepliarsky}, \citenamefont {Stachiotti},\ and\ \citenamefont
  {Migoni}}]{Sepliarsky05p014110}%
  \BibitemOpen
  \bibfield  {author} {\bibinfo {author} {\bibfnamefont {M.}~\bibnamefont
  {Sepliarsky}}, \bibinfo {author} {\bibfnamefont {M.~G.}\ \bibnamefont
  {Stachiotti}}, \ and\ \bibinfo {author} {\bibfnamefont {R.~L.}\ \bibnamefont
  {Migoni}},\ }\href@noop {} {\bibfield  {journal} {\bibinfo  {journal} {Phys.
  Rev. B}\ }\textbf {\bibinfo {volume} {72}},\ \bibinfo {pages} {014110}
  (\bibinfo {year} {2005})}\BibitemShut {NoStop}%
\bibitem [{\citenamefont {Xu}\ \emph {et~al.}(2015)\citenamefont {Xu},
  \citenamefont {Liu}, \citenamefont {Grinberg}, \citenamefont {Karthik},
  \citenamefont {Damodaran}, \citenamefont {Rappe},\ and\ \citenamefont
  {Martin}}]{Xu15p79}%
  \BibitemOpen
  \bibfield  {author} {\bibinfo {author} {\bibfnamefont {R.}~\bibnamefont
  {Xu}}, \bibinfo {author} {\bibfnamefont {S.}~\bibnamefont {Liu}}, \bibinfo
  {author} {\bibfnamefont {I.}~\bibnamefont {Grinberg}}, \bibinfo {author}
  {\bibfnamefont {J.}~\bibnamefont {Karthik}}, \bibinfo {author} {\bibfnamefont
  {A.~R.}\ \bibnamefont {Damodaran}}, \bibinfo {author} {\bibfnamefont {A.~M.}\
  \bibnamefont {Rappe}}, \ and\ \bibinfo {author} {\bibfnamefont {L.~W.}\
  \bibnamefont {Martin}},\ }\href@noop {} {\bibfield  {journal} {\bibinfo
  {journal} {Nat. Mater.}\ }\textbf {\bibinfo {volume} {14}},\ \bibinfo {pages}
  {79} (\bibinfo {year} {2015})}\BibitemShut {NoStop}%
\bibitem [{\citenamefont {Hiroyuki~Takenaka}\ and\ \citenamefont
  {Rappe}(2013)}]{Takenaka13p147602}%
  \BibitemOpen
  \bibfield  {author} {\bibinfo {author} {\bibfnamefont {I.~G.}\ \bibnamefont
  {Hiroyuki~Takenaka}}\ and\ \bibinfo {author} {\bibfnamefont {A.~M.}\
  \bibnamefont {Rappe}},\ }\href@noop {} {\bibfield  {journal} {\bibinfo
  {journal} {Phys. Rev. Lett.}\ }\textbf {\bibinfo {volume} {110}},\ \bibinfo
  {pages} {147602} (\bibinfo {year} {2013})}\BibitemShut {NoStop}%
\bibitem [{\citenamefont {Shin}\ \emph {et~al.}(2005)\citenamefont {Shin},
  \citenamefont {Cooper}, \citenamefont {Grinberg},\ and\ \citenamefont
  {Rappe}}]{Shin05p054104}%
  \BibitemOpen
  \bibfield  {author} {\bibinfo {author} {\bibfnamefont {Y.-H.}\ \bibnamefont
  {Shin}}, \bibinfo {author} {\bibfnamefont {V.~R.}\ \bibnamefont {Cooper}},
  \bibinfo {author} {\bibfnamefont {I.}~\bibnamefont {Grinberg}}, \ and\
  \bibinfo {author} {\bibfnamefont {A.~M.}\ \bibnamefont {Rappe}},\ }\href
  {http://dx.doi.org/10.1103/PhysRevB.71.054104} {\bibfield  {journal}
  {\bibinfo  {journal} {Phys. Rev. B}\ }\textbf {\bibinfo {volume} {71}},\
  \bibinfo {pages} {054104} (\bibinfo {year} {2005})}\BibitemShut {NoStop}%
\bibitem [{\citenamefont {Pertsev}\ and\ \citenamefont
  {Zembilgotov}(1996)}]{Pertsev96p6401}%
  \BibitemOpen
  \bibfield  {author} {\bibinfo {author} {\bibfnamefont {N.~A.}\ \bibnamefont
  {Pertsev}}\ and\ \bibinfo {author} {\bibfnamefont {A.~G.}\ \bibnamefont
  {Zembilgotov}},\ }\href {\doibase 10.1063/1.363659} {\bibfield  {journal}
  {\bibinfo  {journal} {J. Appl. Phys.}\ }\textbf {\bibinfo {volume} {80}},\
  \bibinfo {pages} {6401} (\bibinfo {year} {1996})}\BibitemShut {NoStop}%
\bibitem [{\citenamefont {Wemple}\ \emph {et~al.}(1969)\citenamefont {Wemple},
  \citenamefont {DiDomenico},\ and\ \citenamefont {Jayaraman}}]{Wemple69p547}%
  \BibitemOpen
  \bibfield  {author} {\bibinfo {author} {\bibfnamefont {S.~H.}\ \bibnamefont
  {Wemple}}, \bibinfo {author} {\bibfnamefont {M.}~\bibnamefont {DiDomenico}},
  \ and\ \bibinfo {author} {\bibfnamefont {A.}~\bibnamefont {Jayaraman}},\
  }\href {\doibase 10.1103/physrev.180.547} {\bibfield  {journal} {\bibinfo
  {journal} {Phys. Rev.}\ }\textbf {\bibinfo {volume} {180}},\ \bibinfo {pages}
  {547} (\bibinfo {year} {1969})}\BibitemShut {NoStop}%
\bibitem [{\citenamefont {Boresch}\ \emph {et~al.}(2000)\citenamefont
  {Boresch}, \citenamefont {H\"{o}chtl},\ and\ \citenamefont
  {Steinhauser}}]{Boresch00p8743}%
  \BibitemOpen
  \bibfield  {author} {\bibinfo {author} {\bibfnamefont {S.}~\bibnamefont
  {Boresch}}, \bibinfo {author} {\bibfnamefont {P.}~\bibnamefont {H\"{o}chtl}},
  \ and\ \bibinfo {author} {\bibfnamefont {O.}~\bibnamefont {Steinhauser}},\
  }\href {\doibase 10.1021/jp0008905} {\bibfield  {journal} {\bibinfo
  {journal} {J. Phys. Chem. B}\ }\textbf {\bibinfo {volume} {104}},\ \bibinfo
  {pages} {8743} (\bibinfo {year} {2000})}\BibitemShut {NoStop}%
\bibitem [{\citenamefont {Ponomareva}\ \emph
  {et~al.}(2007{\natexlab{a}})\citenamefont {Ponomareva}, \citenamefont
  {Bellaiche},\ and\ \citenamefont {Resta}}]{Ponomareva07p227601}%
  \BibitemOpen
  \bibfield  {author} {\bibinfo {author} {\bibfnamefont {I.}~\bibnamefont
  {Ponomareva}}, \bibinfo {author} {\bibfnamefont {L.}~\bibnamefont
  {Bellaiche}}, \ and\ \bibinfo {author} {\bibfnamefont {R.}~\bibnamefont
  {Resta}},\ }\href {\doibase 10.1103/PhysRevLett.99.227601} {\bibfield
  {journal} {\bibinfo  {journal} {Phys. Rev. Lett.}\ }\textbf {\bibinfo
  {volume} {99}},\ \bibinfo {pages} {227601} (\bibinfo {year}
  {2007}{\natexlab{a}})}\BibitemShut {NoStop}%
\bibitem [{\citenamefont {Ponomareva}\ \emph
  {et~al.}(2007{\natexlab{b}})\citenamefont {Ponomareva}, \citenamefont
  {Bellaiche},\ and\ \citenamefont {Resta}}]{Ponomareva07p235403}%
  \BibitemOpen
  \bibfield  {author} {\bibinfo {author} {\bibfnamefont {I.}~\bibnamefont
  {Ponomareva}}, \bibinfo {author} {\bibfnamefont {L.}~\bibnamefont
  {Bellaiche}}, \ and\ \bibinfo {author} {\bibfnamefont {R.}~\bibnamefont
  {Resta}},\ }\href {\doibase 10.1103/physrevb.76.235403} {\bibfield  {journal}
  {\bibinfo  {journal} {Phys. Rev. B}\ }\textbf {\bibinfo {volume} {76}},\
  \bibinfo {pages} {235403} (\bibinfo {year} {2007}{\natexlab{b}})}\BibitemShut
  {NoStop}%
\bibitem [{\citenamefont {Ponomareva}\ \emph {et~al.}(2008)\citenamefont
  {Ponomareva}, \citenamefont {Bellaiche}, \citenamefont {Ostapchuk},
  \citenamefont {Hlinka},\ and\ \citenamefont {Petzelt}}]{Ponomareva08p012102}%
  \BibitemOpen
  \bibfield  {author} {\bibinfo {author} {\bibfnamefont {I.}~\bibnamefont
  {Ponomareva}}, \bibinfo {author} {\bibfnamefont {L.}~\bibnamefont
  {Bellaiche}}, \bibinfo {author} {\bibfnamefont {T.}~\bibnamefont
  {Ostapchuk}}, \bibinfo {author} {\bibfnamefont {J.}~\bibnamefont {Hlinka}}, \
  and\ \bibinfo {author} {\bibfnamefont {J.}~\bibnamefont {Petzelt}},\ }\href
  {\doibase 10.1103/PhysRevB.77.012102} {\bibfield  {journal} {\bibinfo
  {journal} {Phys. Rev. B}\ }\textbf {\bibinfo {volume} {77}},\ \bibinfo
  {pages} {012102} (\bibinfo {year} {2008})}\BibitemShut {NoStop}%
\bibitem [{\citenamefont {Yu}\ \emph {et~al.}(2009)\citenamefont {Yu},
  \citenamefont {Ranjan}, \citenamefont {Nardelli},\ and\ \citenamefont
  {Bernholc}}]{Yu09p165432}%
  \BibitemOpen
  \bibfield  {author} {\bibinfo {author} {\bibfnamefont {L.}~\bibnamefont
  {Yu}}, \bibinfo {author} {\bibfnamefont {V.}~\bibnamefont {Ranjan}}, \bibinfo
  {author} {\bibfnamefont {M.~B.}\ \bibnamefont {Nardelli}}, \ and\ \bibinfo
  {author} {\bibfnamefont {J.}~\bibnamefont {Bernholc}},\ }\href {\doibase
  10.1103/PhysRevB.80.165432} {\bibfield  {journal} {\bibinfo  {journal} {Phys.
  Rev. B}\ }\textbf {\bibinfo {volume} {80}},\ \bibinfo {pages} {165432}
  (\bibinfo {year} {2009})}\BibitemShut {NoStop}%
\bibitem [{\citenamefont {Pilania}\ and\ \citenamefont
  {Ramprasad}(2012)}]{Pilania12p7580}%
  \BibitemOpen
  \bibfield  {author} {\bibinfo {author} {\bibfnamefont {G.}~\bibnamefont
  {Pilania}}\ and\ \bibinfo {author} {\bibfnamefont {R.}~\bibnamefont
  {Ramprasad}},\ }\href {\doibase 10.1007/s10853-012-6411-5} {\bibfield
  {journal} {\bibinfo  {journal} {J Mater Sci}\ }\textbf {\bibinfo {volume}
  {47}},\ \bibinfo {pages} {7580} (\bibinfo {year} {2012})}\BibitemShut
  {NoStop}%
\bibitem [{\citenamefont {Foster}\ \emph {et~al.}(1993)\citenamefont {Foster},
  \citenamefont {Li}, \citenamefont {Grimsditch}, \citenamefont {Chan},\ and\
  \citenamefont {Lam}}]{Foster93p10160}%
  \BibitemOpen
  \bibfield  {author} {\bibinfo {author} {\bibfnamefont {C.~M.}\ \bibnamefont
  {Foster}}, \bibinfo {author} {\bibfnamefont {Z.}~\bibnamefont {Li}}, \bibinfo
  {author} {\bibfnamefont {M.}~\bibnamefont {Grimsditch}}, \bibinfo {author}
  {\bibfnamefont {S.-K.}\ \bibnamefont {Chan}}, \ and\ \bibinfo {author}
  {\bibfnamefont {D.~J.}\ \bibnamefont {Lam}},\ }\href {\doibase
  10.1103/PhysRevB.48.10160} {\bibfield  {journal} {\bibinfo  {journal} {Phys.
  Rev. B}\ }\textbf {\bibinfo {volume} {48}},\ \bibinfo {pages} {10160}
  (\bibinfo {year} {1993})}\BibitemShut {NoStop}%
\bibitem [{\citenamefont {Fontana}\ \emph {et~al.}(1991)\citenamefont
  {Fontana}, \citenamefont {Idrissi}, \citenamefont {Kugel},\ and\
  \citenamefont {Wojcik}}]{Fontana91p8695}%
  \BibitemOpen
  \bibfield  {author} {\bibinfo {author} {\bibfnamefont {M.~D.}\ \bibnamefont
  {Fontana}}, \bibinfo {author} {\bibfnamefont {H.}~\bibnamefont {Idrissi}},
  \bibinfo {author} {\bibfnamefont {G.~E.}\ \bibnamefont {Kugel}}, \ and\
  \bibinfo {author} {\bibfnamefont {K.}~\bibnamefont {Wojcik}},\ }\href
  {\doibase 10.1088/0953-8984/3/44/014} {\bibfield  {journal} {\bibinfo
  {journal} {J. Phys.: Condens. Matter}\ }\textbf {\bibinfo {volume} {3}},\
  \bibinfo {pages} {8695} (\bibinfo {year} {1991})}\BibitemShut {NoStop}%
\bibitem [{\citenamefont {Bungaro}\ and\ \citenamefont
  {Rabe}(2004)}]{Bungaro04p184101}%
  \BibitemOpen
  \bibfield  {author} {\bibinfo {author} {\bibfnamefont {C.}~\bibnamefont
  {Bungaro}}\ and\ \bibinfo {author} {\bibfnamefont {K.~M.}\ \bibnamefont
  {Rabe}},\ }\href {\doibase 10.1103/PhysRevB.69.184101} {\bibfield  {journal}
  {\bibinfo  {journal} {Phys. Rev. B}\ }\textbf {\bibinfo {volume} {69}},\
  \bibinfo {pages} {184101} (\bibinfo {year} {2004})}\BibitemShut {NoStop}%
\bibitem [{\citenamefont {Park}\ \emph {et~al.}(2001)\citenamefont {Park},
  \citenamefont {Peterson}, \citenamefont {Jia}, \citenamefont {Lee},
  \citenamefont {Zeng}, \citenamefont {Si},\ and\ \citenamefont
  {Xi}}]{Park01p533}%
  \BibitemOpen
  \bibfield  {author} {\bibinfo {author} {\bibfnamefont {B.~H.}\ \bibnamefont
  {Park}}, \bibinfo {author} {\bibfnamefont {E.~J.}\ \bibnamefont {Peterson}},
  \bibinfo {author} {\bibfnamefont {Q.~X.}\ \bibnamefont {Jia}}, \bibinfo
  {author} {\bibfnamefont {J.}~\bibnamefont {Lee}}, \bibinfo {author}
  {\bibfnamefont {X.}~\bibnamefont {Zeng}}, \bibinfo {author} {\bibfnamefont
  {W.}~\bibnamefont {Si}}, \ and\ \bibinfo {author} {\bibfnamefont {X.~X.}\
  \bibnamefont {Xi}},\ }\href {\doibase 10.1063/1.1340863} {\bibfield
  {journal} {\bibinfo  {journal} {Appl. Phys. Lett.}\ }\textbf {\bibinfo
  {volume} {78}},\ \bibinfo {pages} {533} (\bibinfo {year} {2001})}\BibitemShut
  {NoStop}%
\bibitem [{\citenamefont {Miller}\ and\ \citenamefont
  {Weinreich}(1960)}]{Miller60p1460}%
  \BibitemOpen
  \bibfield  {author} {\bibinfo {author} {\bibfnamefont {R.~C.}\ \bibnamefont
  {Miller}}\ and\ \bibinfo {author} {\bibfnamefont {G.}~\bibnamefont
  {Weinreich}},\ }\href@noop {} {\bibfield  {journal} {\bibinfo  {journal}
  {Phys. Rev.}\ }\textbf {\bibinfo {volume} {117}},\ \bibinfo {pages} {1460}
  (\bibinfo {year} {1960})}\BibitemShut {NoStop}%
\bibitem [{\citenamefont {Tybell}\ \emph {et~al.}(2002)\citenamefont {Tybell},
  \citenamefont {Paruch}, \citenamefont {Giamarchi},\ and\ \citenamefont
  {Triscone}}]{Tybell02p97601}%
  \BibitemOpen
  \bibfield  {author} {\bibinfo {author} {\bibfnamefont {T.}~\bibnamefont
  {Tybell}}, \bibinfo {author} {\bibfnamefont {P.}~\bibnamefont {Paruch}},
  \bibinfo {author} {\bibfnamefont {T.}~\bibnamefont {Giamarchi}}, \ and\
  \bibinfo {author} {\bibfnamefont {J.-M.}\ \bibnamefont {Triscone}},\
  }\href@noop {} {\bibfield  {journal} {\bibinfo  {journal} {Phys. Rev. Lett.}\
  }\textbf {\bibinfo {volume} {89}},\ \bibinfo {pages} {097601} (\bibinfo
  {year} {2002})}\BibitemShut {NoStop}%
\bibitem [{\citenamefont {Mabud}\ and\ \citenamefont
  {Glazer}(1979)}]{Mabud79p49}%
  \BibitemOpen
  \bibfield  {author} {\bibinfo {author} {\bibfnamefont {S.~A.}\ \bibnamefont
  {Mabud}}\ and\ \bibinfo {author} {\bibfnamefont {A.~M.}\ \bibnamefont
  {Glazer}},\ }\href {\doibase 10.1107/s0021889879011754} {\bibfield  {journal}
  {\bibinfo  {journal} {J. Appl. Cryst.}\ }\textbf {\bibinfo {volume} {12}},\
  \bibinfo {pages} {49} (\bibinfo {year} {1979})}\BibitemShut {NoStop}%
\end{thebibliography}
%

\newpage

\begin{table}
\centering
\caption{Structure parameters of tetragonal PbTiO$_3$ optimized with different methods.}
\begin{tabular}{lllllll}
\hline 
\hline
DFT & PSP & $E_c$(Ry) & $E_{\rho}$ (Ry) & $a$ (\AA) & $c$ (\AA) & $c/a$ \\ 
\hline 
LDA & GBRV USPP & 50 & 250 & 3.861 & 4.029 & 1.044 \\ 
 
LDA & GBRV USPP & 60 & 500 & 3.861 & 4.034 & 1.045 \\ 
 
LDA & PSlibrary USPP & 50 & 250 & 3.868 & 4.012 & 1.037 \\ 
 
LDA & PSlibrary USPP & 60 & 500 & 3.861 & 4.034 & 1.045 \\ 

LDA & NC & 50 & • & 3.859 & 3.986 & 1.033 \\ 

LDA & NC & 60 & • & 3.863 & 4.051 & 1.049 \\ 
 
WC & PSlibrary USPP & 50 & 250 & 3.886 & 4.136 & 1.064 \\ 

WC & PSlibrary USPP & 60 & 500 & 3.888 & 4.086 & 1.051 \\ 

PBEsol & GBRV USPP & 50 & 250 & 3.866 & 4.222 & 1.092 \\ 

PBEsol & GBRV USPP & 60 & 500 & 3.866 & 4.243 & 1.098 \\ 

PBEsol & PSlibrary USPP & 50 & 250 & 3.881 & 4.162 & 1.072 \\ 

PBEsol & PSlibrary USPP & 60 & 500 & 3.864 & 4.243 & 1.098 \\ 

PBEsol & NC & 50 & • & 3.888 & 4.106 & 1.056 \\ 

PBEsol & NC & 60 & • & 3.875 & 4.247 & 1.096 \\ 
\hline
experiment & ref~\cite{Mabud79p49} & • & • & 3.90 & 4.15 & 1.064 \\
\hline 
\hline
\end{tabular} 
\end{table}
%experiment & ref~\cite{} & • & • & 3.878 & 4.174 & 1.076 \\

\begin{table}
\centering
\caption{180$^\circ$ DW structure parameters and energies obtained with different methods.}
\begin{tabular}{lcc}
\hline \hline
Model:DFT:PP & ($a,b,c$) & $E_{\rm DW}$(mJ/m$^2$) \\ 
\hline 
I:LDA:GBRV & (3.900, 3.900, 4.150) & 199 \\ 

I:LDA:NC &  (3.900, 3.900, 4.150) & 200 \\ 

I:WC:PSlibrary &  (3.900, 3.900, 4.150) & 166 \\ 

I:PBEsol:GBRV &  (3.900, 3.900, 4.150) & 178 \\ 
 
I:PBEsol:NC &  (3.900, 3.900, 4.150) & 175 \\ 
\hline 
II:LDA:GBRV & (3.861, 3.861, 4.034) & 131 \\ 
 
II:LDA:NC & (3.863, 3.863, 4.051) & 141 \\ 

II:WC:PSlibrary & (3.888, 3.888, 4.086) & 128 \\ 

II:PBEsol:GBRV & (3.864, 3.864, 4.243) & 219 \\ 

II:PBEsol:NC & (3.875, 3.875, 4.247) & 218 \\ 
 
\hline
III:LDA:GBRV & (3.872, 3.870, 3.980) & 102 \\ 

III:LDA:NC & (3.875, 3.871, 3.993) & 108 \\ 

III:WC:PSLibrary & (3.901, 3.899, 4.022) & 95 \\

III:PBEsol:GBRV & (3.896, 3.891, 4.081) & 141 \\ 

III:PBEsol:NC & (3.905, 3.899, 4.098) & 147 \\ 

\hline \hline
\end{tabular} 
\end{table}

\newpage
 
\newpage
 
\newpage

\newpage
 
 \newpage
 
\newpage

\vspace{-4pt}
 
\end{document}